\documentstyle[preprint,eqsecnum,aps,prb]{revtex}

\begin{document}

\title{Ab Initio Study of the Ferroelectric Transition in Cubic Pb$_3$GeTe$_4$}

\author{Eric Cockayne and Karin M. Rabe}

\address{Department of Applied Physics,
Yale University,
P.O. Box 208284,
New Haven, CT 06520-8284}
\date{\today}
\maketitle

\begin{abstract}

	In the substitutionally disordered narrow-gap semiconductor
Pb$_{1-x}$Ge$_x$Te, a finite-temperature cubic-rhombohedral transition
appears above a critical concentration $x \approx 0.005$.  
As a first step towards a
first-principles investigation of this transition in the disordered
system, a
(hypothetical)
ordered cubic Pb$_3$GeTe$_4$ supercell is studied.  First principles 
density-functional calculations of total energies and linear response functions
are performed using the conjugate-gradients method with ab initio
pseudopotentials and a plane-wave basis set.  Unstable modes in
Pb$_3$GeTe$_4$ are found, dominated by off-centering of the Ge ions coupled with
displacements of their neighboring Te ions.
A model
Hamiltonian for this system is constructed using the
lattice Wannier function formalism.
The parameters for this Hamiltonian are determined from first principles.
The equilibrium thermodynamics of the model system is
studied via Metropolis
Monte Carlo simulations.  The calculated transition temperature,
$T_c$, is approximately 620K for the cubic 
Pb$_3$GeTe$_4$ model,
compared to the experimental value of $T_c 
\approx 350K$ for disordered Pb$_{0.75}$Ge$_{0.25}$Te.  
Generalization of this
analysis to the disordered Pb$_{1-x}$Ge$_x$Te system is discussed.

\end{abstract}

\pacs{PACS Numbers: 77.80.Bh, 64.60.Cn, 77.84.Bw, 63.20.Dj}

\section{Introduction}

	Numerous substitutionally disordered ferroelectrics
exhibit structural phase
transitions in  which the nature of the phase transition
changes as a function of composition.  Examples include
Ba$_{1-x}$Sr$_x$TiO$_3$\cite{Lem96},
PbZr$_{1-x}$Ti$_x$O$_3$\cite{Shi52,Fes86}, and
Pb$_{1-x}$Ge$_x$Te\cite{Hoh72}.  
Pb$_{1-x}$Ge$_x$Te 
is an ideal system for the development of
first-principles methods to investigate the effect of 
substitution on structural phase transitions for several
reasons. 
First, the unit cells of the endpoint compounds are small.  
Pure PbTe and GeTe have only two atoms and ten
valence electrons per unit cell, as opposed to five atoms and 
24 or more valence electrons per unit cell in the perovskite oxide
ferroelectrics.
Secondly, the phase diagram in Pb$_{1-x}$Ge$_x$Te is rather simple.\cite{Hoh72}
Pure PbTe has a rocksalt structure that is stable to zero temperature.
Pure GeTe also has the rocksalt structure at high temperature but
undergoes a structural phase transition, indicated in Figure {\ref{gete.ps}}, 
to a rhombohedral phase at
a critical temperature $T_c \approx$ 670K.\cite{Sch53}  As the Ge concentration
$x$ decreases from 1 to 0, $T_c$ for a cubic-rhombohedral
phase transition decreases smoothly to zero
at $x \approx 0.005$.\cite{Hoh72,Tak79}
Finally, there are existing
phenomenological and empirical models for this 
transition.\cite{Log77,Kat80,Yar81,Isl87}
These models focus on the role of Ge off-centering\cite{Log77,Kat80} 
in the phase transition.
Off-centering is observed in a number of compounds
where there is a mismatch between the ionic radii of
two species that statistically occupy the same
kind of site.\cite{Bar75}
There is direct experimental evidence for off-centering in
Pb$_{1-x}$Ge$_x$Te.  The extended X-ray-absorption fine-structure
measurements of Islam and Bunker\cite{Isl87} show two
peaks in the distribution function for Ge-Te distances
both above and below $T_c$,
as opposed to the single peak that would be seen if
the Ge atoms were located at the centers of the octahedra
formed by their first-neighbor Te atoms.
The phenomenological and empirical models previously
considered provide a reference for comparison with the model
derived from first principles here.

	Model Hamiltonians based on ab-initio calculations have been
successfully constructed for a number of stoichiometric ferroelectrics
and related materials,  
including 
GeTe\cite{Rab87}, PbTiO$_3$\cite{Rab95pt}, 
PbZrO$_3$\cite{WaghRab}, BaTiO$_3$\cite{Zho94bt,Vanderbilt},
SrTiO$_3$\cite{Zho96}
and KNbO$_3$\cite{Kra97}. 
In constructing these models,
the high-symmetry average structure of the high-temperature phase is 
chosen as a reference
structure.  Normal mode dispersion relations and eigenfunctions are
calculated to determine the unstable modes.
The normal mode branches containing the unstable modes form
a basis for the ionic displacement subspace determining, by projection, an  
effective
Hamiltonian.  Via a linear transformation, a localized
lattice Wannier function\cite{Rab95lwf} (LWF) basis is found.
The energy is expanded in powers
of the lattice Wannier function coordinates and strain 
and the coefficients 
in the expansion
are determined from a set of ab-initio calculations.

	In a substitutionally disordered system, it is 
not enough to include only the  
lattice degrees of freedom.
Configurational entropy also plays an important role; the partition  
function, from which all
thermodynamic properties can be obtained, includes a sum 
over all possible configurations.  Different {\it fixed}
configurations will undergo structural phase transitions
at different temperatures.
Furthermore, the nature of the
configurations whose properties  
dominate in the thermodynamic limit will itself depend on temperature.
However, it is not necessary to investigate all possible
configurations in order to model a phase transition in a disordered system. 
In fact, if a model can be formulated in terms of intersite interactions that
decay rapidly with distance to some asymptotic form, only a small number 
of configurations need to be explicitly calculated from first principles.

	In this work, we investigate a single
Pb$_{1-x}$Ge$_x$Te configuration: the {\it ordered} 8 atom Pb$_3$GeTe$_4$ 
cubic cell 
($x = 0.25$), shown in Figure {\ref{structure.ps}}.  This configuration was  
chosen for several reasons: (1) it is toward
the Pb-rich end of the phase diagram, which is appropriate for studying
how adding a small concentration of Ge to PbTe ``turns on"
a phase transition at zero temperature, (2) it contains the minimum
number of atoms per cell (eight) for any $x \le 0.25$ and (3) among the
eight atom cells with $x = 0.25$, it has the highest
point symmetry.  The small size and high symmetry of the unit cell for
cubic Pb$_3$GeTe$_4$ makes {\it ab initio} calculations relatively 
computationally inexpensive.  Our
calculation of $T_c$ for this system provides a benchmark for comparison
with $T_c$ for other configurations and for
the ensemble average at the same composition.

	This paper is organized as follows.  Section II describes the 
general principles governing the 
construction of model Hamiltonians and the lattice Wannier
function method for determining the effective Hamiltonian subspace.  
In Section III, these
methods are applied to the specific case of cubic Pb$_3$GeTe$_4$ and
a model Hamiltonian is obtained.  The model Hamiltonian
is used in a classical Monte Carlo simulation in Section IV
to obtain the transition temperature and order of the cubic-
rhombohedral phase transition in this system.  The results are
further discussed in Section V, including their implications for the
substitutionally disordered Pb$_{1-x}$Ge$_x$Te system. Conclusions
are given in Section VI.

\section{Construction of model Hamiltonians}

	At the level of the Born-Oppenheimer approximation for
electronic energy,
the classical partition function $Z$ for a disordered
system on a fixed lattice is
\begin{equation}
\label{eqn21}
Z = \sum_{\{ \sigma_j \}} \int d \{ {\bf u}_{j} \}
\int d \{ \dot{\bf u}_{j} \}
\int d {\bf e}~{\rm exp} [ (- \beta E( \{ \sigma_j \},
\{ {\bf u}_{j} \},
\{ \dot{\bf u}_{j} \},
 {\bf e} ) ],
\end{equation}
where $\{ \sigma_j \}$ represents the chemical configuration,
{\it i.e.} which
type of ion occupies each site $j$, ${\bf e}$ is the
homogeneous strain tensor, $\{ {\bf u}_{j} \}$ is the set of ionic
displacements defined with respect to some reference 
structure for the given chemical configuration and
strain, and $\{ \dot{\bf u}_{j} \}$ is the set of ionic
velocities.  If, in addition, the ionic motion is treated
classically, the integral over 
$\{ \dot{\bf u}_{j} \}$ leads to a trivial 
${\bf u}$-independent factor, and the partition function
can be rewritten as
\begin{equation}
\label{eqn21b}
Z \propto \sum_{\{ \sigma_j \}} \int d \{ {\bf u}_{j} \}
\int d {\bf e}~{\rm exp} [ (- \beta E( \{ \sigma_j \},
\{ {\bf u}_{j} \},
 {\bf e} ) ],
\end{equation}

	The partition function {\ref{eqn21b}} can be rewritten
$Z \propto \sum_{\{ \sigma_j \}} Z( \{ \sigma_j \})$,
where $Z( \{ \sigma_j \})$ is the partition function
for the ensemble corresponding to the single chemical
configuration $\{ \sigma_j \}$.  In what follows, we
consider only one chemical configuration, so the
configuration label $\sigma_j$ will be dropped.  The 
partition function now simplifies to
\begin{equation}
Z \sim \int d \{ {\bf u}_{j} \} 
\int d {\bf e}~{\rm exp} [ (- \beta E( \{ {\bf u}_{j} \},
 {\bf e} ) ].
\end{equation}  
In general, $E( \{ {\bf u}_{j} \},{\bf e} )$ can be
expanded as a Taylor series in powers of ${\bf u}_{j}$
and ${\bf e}$.
If the reference structure is at an energy extremum,
then the linear terms in the expansion vanish:
\begin{equation}
\label{eqn23}
E(\{{\bf u}_i\}, {\bf e}) = E^{(0)} +
E^{(2)} (\{{\bf u}_j\}, {\bf e}) +
E^{(3)} (\{{\bf u}_j\}, {\bf e}) + \dots. 
\end{equation}
The harmonic term, $E^{(2)} (\{{\bf u}_j\}, {\bf e})$
is the sum of a lattice term 
\begin{equation}
\label{eqn23a}
\sum_{ij\alpha\beta} F_{ij\alpha\beta}~u_{i\alpha}~u_{j\beta},
\end{equation}
where $F$ is the force constant matrix,
a strain term
\begin{equation}
\label{eqn23b}
\sum_{\alpha\beta\gamma\delta}~
c_{\alpha\beta\gamma\delta}~
e_{\alpha\beta}~e_{\gamma\delta},
\end{equation}
where $c$ is the elastic constant tensor,
and a strain-coupling term
\begin{equation}
\label{eqn23c}
\sum_{i\alpha\beta\gamma}~g_{i\alpha\beta\gamma}~u_{i\alpha}~
e_{\beta\gamma},
\end{equation}
where $g$ is the strain-coupling tensor.
By the usual change of variable to normal mode variable,
$\{ {\bf u}_j \} = \sum_{\nu} a_{\nu} \epsilon_{\nu}$,
the harmonic lattice term is 
reduced to a single sum:
\begin{equation}
\sum_{ij\alpha\beta} F_{ij\alpha\beta}~u_{i\alpha}~u_{j\beta} =
{1\over{2}} \sum_{\nu} m_{\nu} \omega_{\nu}^2 a_{\nu}^2,
\end{equation}
where $a_{\nu}$ is the amplitude,
$\epsilon_{\nu}$ the (normalized) ionic displacement
pattern, $\omega_{\nu}$
the frequency, and $m_{\nu}$ the mode effective mass
for normal mode $\nu$.

In what follows, it is assumed that the fixed configuration
is periodic.  Then the normal mode label $\nu$ can be
replaced by ${\bf k} i \alpha$, where ${\bf k}$ is the
wavevector, $i$ the symmetry-invariant subspace and 
$\alpha$ the branch of the normal mode dispersion
curves on which the mode lies.  For example,
in a structure with two atoms per unit cell, $i = 1$
represents the acoustic modes, $i = 2$ the optical modes,
and in each subspace $\alpha = 1$ labels the longitudinal
and $\alpha = 2,3$ labels the 
two transverse branches.

If the extremum or reference structure is a saddle point
in the energy surface rather than a minimum, some of the
normal modes will be unstable ($\omega_{{\bf k} i \alpha}^2 < 0$).
As described in detail by Rabe and Waghmare\cite{Rab92,Rab95lwf},
the model Hamiltonian approach to structural phase transitions
associated with soft phonons aims to reduce the ionic degrees 
of freedom to those of the
``effective Hamiltonian" subspace that contains
the important anharmonic terms in Eq. {\ref{eqn23}}.
All normal mode subspaces $i$ that contain
unstable modes must be included in the effective
Hamiltonian subspace.
An expansion of the total energy 
in these degrees in freedom
requires higher-order terms in
order to stabilize the structure at a finite distortion.  
As the simplest approximation, terms are
kept to only harmonic order in the subspace
complementary to the effective Hamiltonian subspace and higher 
order 
mixing of the effective Hamiltonian subspace with the complementary
subspace is neglected.  
This approximation 
is valid to the extent that the magnitude of the neglected higher
order terms is small for all ionic displacement patterns 
that contribute significantly to $Z$, {\it i.e.} those where 
$E(\{{\bf u}_j\}, {\bf e})$ is small. 
Integration over the complementary subspace
gives a structure-independent contribution to $Z$ which can be 
neglected.  The remaining terms in $E$ give an effective
Hamiltonian ${\cal H}_{eff}$.  Integration of ${\cal H}_{eff}$ over
the effective Hamiltonian subspace is sufficient to reproduce
the dependence of the structure on temperature.

	  A change in variables allows the basis of the effective Hamiltonian
subspace
to be converted into one where the displacements are localized, {\it i.e}
a lattice Wannier function (LWF) basis ${\bf w}_{\bf R}$\cite{Rab95lwf}.
This procedure is analogous to that which 
des Cloizeaux\cite{des64} gave for determining {\it electronic}
Wannier functions for a multidimensional subspace; 
except that the periodic
functions are now ionic displacement patterns.\cite{Rab95lwf}
For simplicity, we assume that the effective Hamiltonian subspace
consists of only a single normal mode subspace $i = 1$, drop the
subscript $i$, and let $\epsilon_{{\bf k} \alpha}$ label the normal
modes in this subspace.
The LWF is given in terms of the normal modes by
\begin{equation}
\label{eqn29a}
w_{\beta {\bf R}}  = {V \over {(2 \pi)^3}} \int_{BZ} d{\bf k} (\sum_{\alpha} 
C_{\alpha\beta} ({\bf k}) \epsilon_{{\bf k} \alpha} 
e^{i (\phi_{ {\bf k} \alpha} - {\bf k} \cdot {\bf R})}).
\end{equation}
The subscript $\beta$ labels the components of the LWF.
The number of components of ${\bf w}_R$ is equal to the number of
branches in the effective Hamiltonian normal mode subspace.
The matrix $C ({\bf k})$ is included so that the polarizations of the
different normal mode branches are correctly related to the LWF components
across the Brillouin zone.
The phases $\phi_{ {\bf k} \alpha}$ are very important.  They
are chosen to make the ionic displacement pattern corresponding
to ${\bf w}$ both as local as possible and with as high a symmetry
as possible.
By construction, the LWF has translational symmetry:
$T_{{\bf R}^{\prime}} (w_{\beta{\bf R}}) = 
w_{\beta{\bf R} + {\bf R}^{\prime}}$, where ${\bf R}^{\prime}$
is any lattice vector.

The normal modes are given in terms of the LWF basis
via the inverse transform
\begin{equation}
\label{eqn25}
[ e^{i \phi_{{\bf k} \alpha} } ]  \epsilon_{{\bf k} \alpha} = 
\sum_{{\bf R} \beta} (C^{-1})_{\alpha\beta} w_{\beta {\bf R}} 
e^{i {\bf k} \cdot {\bf R} }.
\end{equation}
Any displacement pattern ${\bf u}_j$ in the effective Hamiltonian 
subspace can be written as
\begin{equation}
{\bf u}_j  = \sum_{{\bf k} \alpha} a_{{\bf k} \alpha} 
\epsilon_{{\bf k} \alpha}.
\end{equation}
Using Eq. {\ref{eqn25}}, and changing the order of the summation, we obtain
\begin{equation}
\label{eqn28}
{\bf u}_j  = \sum_{{\bf R} \beta} w_{\beta {\bf R}}
[\sum_{{\bf k} \alpha} (C^{-1})_{\alpha\beta} a_{{\bf k} \alpha} 
e^{i ({\bf k} \cdot {\bf R} - \phi_{{\bf k} \alpha}) }]
= \sum_{{\bf R} \beta} w_{\beta {\bf R}} \xi_{\beta {\bf R}}.
\end{equation}  
The new vector parameter ${\vec \xi}_ {\bf R}$ gives
the displacement field via Eq. {\ref{eqn28}} and provides the basis for a 
spin-like model of the system.

\section{Model Hamiltonian Construction for {\rm Pb$_3$GeTe$_4$}}

    In this section, we describe the construction of the effective 
Hamiltonian  
subspace
and the explicit expression for the model Hamiltonian for Pb$_3$GeTe$_4$ from 
first-principles density-functional calculations within the local-density  
approximation (LDA).
Bachelet, Hammann and Schl\"uter
pseudopotentials\cite{Bac82} were used for each species in the 
two-projector Kleinman-Bylander
form.\cite{Kle82,Blo90}  For Pb and Te, the $d$ potential was taken as 
local; for Ge,
a $d$ local potential was found to give spurious low-energy ``ghost"  
states\cite{Gon90,Gon91}
and so a $p$ local potential was chosen.
Total energy calculations were done using the
CASTEP~2.1 program of Payne {\it et al.}\cite{CAS91}
CASTEP~2.1 performs conjugate gradients minimization\cite{Pay92} of the
LDA density functional total energy in the Kohn-Sham formalism.\cite{Koh65}
The exchange-correlation energy was Perdew and Zunger's 
parameterization\cite{Per81}
of the Ceperley-Alder values\cite{Cep78} for the uniform electron gas.
The calculations were performed for a 8-atom unit cell with a 300 eV
cutoff for the plane wave basis set (about 3000 plane waves per k-point), 
a 4 by 4 by 4 Monkhorst-Pack\cite{Mon76}
k-point set, and a 36 by 36 by 36 real-space grid for the charge density. 
Force constant matrices at the high-symmetry ${\bf q} \ne 0$ points in the  
Brillouin
zone and Born effective charges were computed using density-functional  
perturbation theory\cite{Bar87,Gia91} in the variational 
formulation.\cite{Gon92}  
At ${\bf q} = 0$, the force constant matrix was computed using 
Hellmann-Feynman  forces, as described in more detail below. Pulay  
corrections\cite{Rig95} 
were added to the total energy results as described in the Appendix.

	We began with Pb$_3$GeTe$_4$ in a high-symmetry reference
structure.  The structure,
shown in Figure {\ref{structure.ps}},
is produced by replacing a cubic superlattice of
Pb ions in the PbTe rocksalt structure with Ge ions.
There are eight atoms per unit
cell.  The structure has full cubic symmetry (space group Pm3m).  
The Ge occupies the cell corner (Wyckoff position 1(a)) and the 
three Pb's occupy the face centers (3(c)).  The four Te's occupy 
two crystallographically distinct positions: the edge centers (3(d)) 
and the cube center (1(b)).
Although this structure is not the minimum-energy structure,
the forces on all ions are zero by symmetry, and
the problem of relaxation can thus be neglected (note this
is not true for the general case of partial Pb $\rightarrow$ 
Ge substitution).  Furthermore, the Pb$_3$GeTe$_4$
structure with the lattice constant $a$ which minimizes the total energy 
is at an extremum of the
$E(\{{\bf u}_i\}, {\bf e})$
energy surface, so that it can be used as a reference structure in 
the analysis of Section II.  We found $a$ = 6.275~\AA~for
cubic Pb$_3$GeTe$_4$.
Although there is no experimental cubic Pb$_3$GeTe$_4$ sample with which
this result can be prepared, we can estimate the ``experimental"
lattice parameter in two different ways as follows.
First, we extrapolate the experimental lattice parameters for PbTe
at room temperature\cite{Lug65} and for GeTe at the cubic-rhombohedral
phase transition temperature\cite{Zhu67} to zero temperature via
the thermal expansion coefficient of PbTe at room temperature\cite{Lug65}
and then average the results assuming Vegard's law to 
obtain $a =$ 6.307(2)~\AA.  Second, we estimate the experimental 
lattice parameter for disordered
Pb$_{0.75}$Ge$_{0.25}$Te at room temperature and the thermal expansion
coefficient for nearby compositions via the figures in Reference
\onlinecite{Hoh72} and then extrapolate to zero temperature to
obtain $a =$ 6.311(1)~\AA.
The two estimates are in good agreement.  The LDA lattice
parameter is about 0.5\% lower than the ``experimental" value.
This magnitude of lattice constant underestimate is 
typical of LDA calculations.\cite{Fil94}

	Next, we identified the lattice instabilities of the 
reference structure.
The force constants at ${\bf q} = 0$ were found
by displacing each ion in turn and then calculating
the Hellmann-Feynman forces.
Anharmonic effects were found to be significant in
Pb$_3$GeTe$_4$ and thus it was necessary to calculate the forces 
for small (0.001 to 0.01~\AA) displacements and extrapolate 
the force constant values to zero displacement in order to obtain high
precision.  For comparison, we also calculated the force
constant matrix at ${\bf q} = 0$ using the density-functional
perturbation theory method.
The latter results, however, violated the acoustic
sum rule by as much as 0.1 eV/\AA$^2$, whereas the maximum
violation of the acoustic sum rule for the extrapolated
Hellmann-Feynman forces was only $10^{-4}$ eV/\AA$^2$.
The discrepancy between the force constant matrix
elements found by the two methods 
was largest for the diagonal terms.  In order to avoid
significant acoustic sum rule violation corrections,
at ${\bf q} = 0$ we used the extrapolated Hellmann-Feynman forces.

     Diagonalization of the resulting ${\bf q} = 0$ dynamical
matrix yielded a single unstable mode 
at $\Gamma$ (see Table \ref{freq.tbl}),
with symmetry $\Gamma_{15}$.   The fact that it is strongly
dominated by Ge motion led us to choose a single LWF centered
on Ge with $\Gamma_{15}$ (vector) symmetry.
To construct the effective Hamiltonian subspace, it 
is therefore necessary only  
to
investigate those ${\bf q} \not= 0$
normal modes compatible with this choice of LWF symmetry.
We performed ${\bf q} \ne 0$ linear-response calculations to calculate
the normal mode frequencies and eigenvectors for 
all such modes at the BZ points 
$\Gamma$, $R$,
$X$ and $M$ (Table \ref{freq.tbl}).
For each label\cite{Bas75} $X_1^{\prime},X_5^{\prime}$, $M_1^{\prime}$,
$M_5^{\prime}$ and $R_{15}$, the lowest energy mode also has a
large component of Ge motion
(see Tables \ref{freq.tbl} and \ref{mode.tbl}), and is therefore 
easily identified
as a mode belonging to the effective Hamiltonian subspace.

	Substitution of the first-principles results
for the ionic displacements of the selected normal modes
into Eq. {\ref{eqn25}} leads to a set of linear equations, which
can be solved to obtain the LWF.  To obtain
the exact LWF, which in general involves nonzero ionic displacements 
out to infinite distance, it would be necessary to include an infinite 
number  of independent normal modes in the analysis.  
We expect, however, that the ionic displacements corresponding
to the lattice Wannier function decrease 
rapidly with distance, in analogy to the behavior  
of electronic Wannier functions\cite{des64} and of the
lattice Wannier function for PbTiO$_3$.\cite{Rab95pt}.
A good approximation to such a LWF can be obtained by
setting all ionic 
displacements equal to zero outside of some finite region.  
The remaining finite set of ionic displacements can be determined from a
finite set of normal modes via Eq.~\ref{eqn29a}.
For cubic Pb$_3$GeTe$_4$, we used all the modes of
Table \ref{mode.tbl}. 
All modes were normalized so that the sum of the squared ionic 
displacements 
was 1~\AA$^2$ per primitive cell.
For these high symmetry points, the normal modes can be chosen so that
Ge motion is always strictly along a Cartesian direction and thus the
matrix $C$ in Eq. {\ref{eqn29a}} is the unit matrix.
The phases $\phi_{{\bf k}\alpha}$ are already incorporated into 
Table \ref{mode.tbl} and are chosen so that Ge motion is always in the
same direction.
The included modes yielded a set of 21 independent linear equations and
therefore allowed the 21 independent components of ionic displacement
shown in 
Table {\ref{lwf.tbl}} to be determined.
The displacements corresponding to the approximate LWF 
involve all ions out to $(\sqrt{5}/2) a$ from the central 
Ge ion and some ions as far as $1.5 a$ away.

	As can be seen from Table {\ref{lwf.tbl}} 
and Figure {\ref{lwf.ps}}, the LWF  
satisfies the locality 
assumption very well, with
first-near neighbor Te displacements only 0.27 times the central Ge 
displacement, second near-neighbor Pb displacements up to 0.14 times
the Ge displacement and all further-neighbor displacements less than 
0.05 times the central Ge displacement.  
Note that although the LWF component shown transforms as the
$z$ component of a vector, individual ionic displacements, such
as the Pb displacements shown in Figure~\ref{lwf.ps}, can be
in directions other than $z$.
Henceforth, we use units 
in which a dimensionless 
local distortion amplitude of 1 in the 
$z$ direction corresponds to the displacements shown in
Table {\ref{lwf.tbl}}.

	To summarize, the
structure-dependent part of the partition function
for Pb$_3$GeTe$_4$ should depend only on a subspace of the
full ionic displacement space, which can be used to
generate an effective Hamiltonian.  So far, we
have found a good approximation to 
${\bf w}_{\bf R}$, the lattice Wannier function basis for
the effective Hamiltonian subspace, which includes
the unstable normal modes.  We have a
spin-like representation in which there is a one-to-one correspondence
between ionic displacement patterns in the effective Hamiltonian 
subspace and the magnitudes and directions of a set of vectors
located on each Ge ion.  We will now
describe the model Hamiltonian for Pb$_3$GeTe$_4$, obtained
by (1) expanding the effective Hamiltonian
energy per unit cell
${\cal H}_{eff}(\{ \vec\xi_i \},{\bf e})/N$ in powers
of $\vec \xi_i$ and strain {\bf e}, (2) truncating this expansion,
and (3) determining the coefficient for each term from first-principles  
calculations.  

Previous ab initio studies on perovskite ferroelectrics such as
BaTiO$_3$\cite{Zho94bt} and PbTiO$_3$\cite{Rab95pt} 
have led to models of the same nature as that presented here, 
namely a set of interacting vector spins
sitting on Wyckoff positions of full cubic symmetry
for a cubic lattice with underlying space group $Pm3m$.
For Pb$_3$GeTe$_4$, we used the same truncations in the expansion of
${\cal H}_{eff}$ as was used 
for the PbTiO$_3$ model.\cite{Rab95pt}  We verified that the terms 
retained were necessary and sufficient to accurately fit the
corresponding {\it ab initio} results.
Various contributions to
the model Hamiltonian will now be considered in turn.

The total energy contains a structure-independent constant term
\begin{equation}
\label{eqn31}
U_0/N
\end{equation}
giving the energy of the high symmetry
Pb$_3$GeTe$_4$ reference structure ($\vec \xi_i = {\bf e} = 0$).  
In the case of disordered systems,
differences in reference structure 
energy are important for comparing different chemical configurations.  
The value of $U_0/N$ for cubic Pb$_3$GeTe$_4$ is given in the Appendix.
Since only one cation configuration is considered in this work,
the value of $U_0/N$ does not affect subsequent results. 

	The lowest order terms which depend on the 
local distortions $\{\vec \xi_i\}$  
are harmonic, with a local contribution
$U_{lh} = \sum_i A |\vec\xi_i|^2$
and an intersite pair contribution 
$U_{ph} = \sum_{i,j}a_{i j \alpha \beta}  
\xi_{i\alpha}\xi_{j\beta}$.
At long range, we expect the intersite pair interaction
to approach the dipole-dipole form, while at
shorter range, there should be significant corrections due to higher 
order multipole effects, induced
charge redistributions and direct overlap of
the ionic displacement patterns.
We thus
split
$U_{ph}$ 
into the sum of a long-range dipole-dipole 
term,
$U_{dd}$, and a short range ``correction"
term, $U_{sr}$.

The energy of a system of
dipoles of moment $\vec\mu$ in a medium of
electronic dielectric constant $\epsilon_{\infty}$ is given
by:\cite{Pic70}
\begin{equation}
\label{eqn32}
U_{dd}(\{ \vec\mu_i \})/N =
{1\over N}\sum_i\sum_{\vec d}{1 \over
\epsilon_\infty}
{(\vec\mu_i\cdot\vec\mu_{i+\vec d}-
3(\vec\mu_i \cdot \hat d)(\vec\mu_{i+\vec d}\cdot \hat d))\over 
|\vec d|^{3}}.
\end{equation}
The sum over $\vec d$ is a sum over all interdipole separations,
$\vec\mu_i$ is the dipole moment at site $i$, and $\vec\mu_{i +\vec d}$ 
is the dipole moment of the dipole at distance $\vec d$ from the dipole
at site $i$.

	To make use of Eq.~{\ref{eqn32}}, we need to know the
dipole moments $\vec\mu_i$ corresponding to local
distortions $\vec\xi_i$. 
Since the dipole moment at 
$\vec\xi = 0$ is zero via centrosymmetry, the dipole
moment for nonzero $\vec\xi$ is simply the sum of the 
polarizations $\vec P_j$ induced by individual ionic displacements  
$\vec u_j$  specified by the given $\vec\xi_i$:
\begin{equation}
\vec\mu = \sum_j \vec P_j (\vec u_j(\vec\xi_i)).
\end{equation}
For small displacements, the $\vec P_j$ are determined  
by the Born effective charge tensor for the ion $j$, defined
as the
differential change in total polarization
due to displacement of that ion:
\begin{equation}
Z^{\star}_{j\alpha\beta} \equiv
{{\partial P_{\alpha}}\over{\partial u_{j \beta}}}.
\end{equation}
Symmetry considerations reduce the number of independent
terms in the Born effective charge tensors for
Pb$_3$GeTe$_4$.
Though our Pb$_3$GeTe$_4$ structure has cubic symmetry,
not all of the ions sit at centers of full cubic symmetry.
The Ge and Te(1) ions sit at centers of cubic 
symmetry, while the Pb
and Te(2) ions fill the two Wyckoff positions of
multiplicity three, each
of which is a center of tetragonal symmetry.  The Born effective
charge tensor is still diagonal under tetragonal symmetry, but
the axial and perpendicular values are different.
Thus, there are 6 independent Born effective charge tensor
components 
in the Pb$_3$GeTe$_4$ cell, one each for Ge and Te(1) and two each 
for Pb and Te(2).
The Born effective charge tensors, obtained from
linear response calculations for
cubic Pb$_3$GeTe$_4$ at the equilibrium
LDA lattice constant, are shown in Table {\ref{born.tbl}}.
The sum of the Born effective charges
$\sum_{j}^{{\rm unit cell}} ({1\over{3}} Z^{\star}_{j \parallel} +
{2\over{3}} Z^{\star}_{j \perp})$
as determined via linear response calculations
was $-0.36$.  This violation of charge neutrality
is a consequence of finite k-point sampling.\cite{Gon97}  We
corrected for it\cite{Gon97} by adding the same constant 
0.045 to all diagonal Born effective charge tensor components.
The diagonal cation (Te) effective charge components are in the 
range $+6$ to $+8$ ($-6$ to $-8$).  These effective charges are
anomalously large
compared to the formal valence charge states, 
but consistent in magnitude with those found 
experimentally\cite{Bur70}
and theoretically\cite{Lit79} for binary group IV tellurides.  

	Since the $u_j$ are proportional to 
$\vec \xi_i$ 
(Eq. {\ref{eqn28}}), the  
dipole moment $\vec\mu$ can be written simply as 
\begin{equation}
\label{eqn35}
\vec\mu_i = \sum_j {\bf Z^{\star}_j} {\bf u}_j (\vec\xi_i) =
\overline Z^{\star} (\vec\xi_i),
\end{equation}
where symmetry leads to a scalar 
mode effective charge parameter
$\overline Z^{\star}$, which for
Pb$_3$GeTe$_4$ has the value 12.10 e\AA.
Eq. {\ref{eqn32}} can then be written in terms of $\vec\xi$ as follows:
\begin{equation}
\label{eqn36}
U_{dd}(\{ \vec\xi_i \})/N =
{1\over N}\sum_i\sum_{\vec d}{(\overline Z^{\star})^2 \over
\epsilon_\infty}
{(\vec\xi_i\cdot\vec\xi_{i+\vec d}-
3(\vec\xi_i \cdot \hat d)(\vec\xi_{i+\vec d}\cdot \hat d))\over 
|\vec d|^{3}}
\end{equation}

	The other material specific parameter 
needed in Eq. {\ref{eqn32}} is $\epsilon_{\infty}$. By  
symmetry, the dielectric tensor is isotropic in cubic
Pb$_3$GeTe$_4$.  
Via 
linear response methods, we calculated
$\epsilon_{\infty}$ = 46.7.
The  experimental value for PbTe is
33-34 at room temperature\cite{Bur70,Mur79} and 40 at zero
temperature.\cite{Mur79}
For GeTe, the measured value\cite{Bur70} at room temperature is 36.
For Pb$_{1-x}$Ge$_x$Te, $\epsilon_{\infty}$ 
measurements are available for $x = 0.07$,
for which the values obtained are about 39 at room temperature,
41 for the cubic phase just above the phase transition and
40 at zero temperature.\cite{Mur79}
Our value for $\epsilon_{\infty}$ is somewhat higher than the value
obtained from interpolation of the experimental values,
consistent with the general tendency of LDA to overestimate
dielectric constants.\cite{Lev89}
 
	Next, consider $U_{sr}$, the short-range correction term in
the intersite pair interaction.  
The force constant matrix
elements determined in a series of ${\bf q} \ne 0$ 
linear
response calculations, along with the corresponding
ionic displacement fields, allow the total harmonic 
energy $U_{lh}+U_{ph}$ to be determined.
These are given  in Table {\ref{harm.tbl}} for the various high-symmetry 
${\bf q}$ points
used in determining the LWF.  
For each normal mode used to construct the LWF, 
we calculated the dipole-dipole interaction energy $U_{dd}$ via Eq.  
{\ref{eqn36}}.
We then subtracted the dipole part from
the total harmonic energy to obtain $U_{sr}(\{ \vec\xi_i \})$.

	As in the case of the LWF itself, the
information obtained from a finite set of independent
$q$ points limits the number of interaction 
coefficients that can be determined.
We include here all symmetry-adapted couplings
to third neighbors in the cubic lattice (or sixth neighbor
cations). There are two independent coefficients at first neighbor
\begin{equation}
\label{eqn37}
U_{sr}(\{ \vec\xi_i \})/N =
{1\over N}\sum_i\sum_{\hat d=nn1}[
b_L(\vec\xi_i\cdot \hat d)(\vec\xi_{i+\hat d}\cdot \hat d)
+b_T(\vec\xi_i\cdot\vec\xi_{i+\hat d}
-(\vec\xi_i\cdot \hat d)(\vec\xi_{i+\hat d}\cdot \hat d))],
\end{equation}
three at second neighbor
\begin{eqnarray}
& + &{1\over N}\sum_i\sum_{\hat d=nn2}[
 d_L(\vec\xi_i\cdot \hat d)(\vec\xi_{i+\hat d}\cdot \hat d)
 + d_{T1}(\vec\xi_i\cdot \hat d_1)(\vec\xi_{i+\hat d}\cdot \hat d_1) 
\nonumber \\
& + & d_{T2}(\vec\xi_i\cdot \hat d_2)(\vec\xi_{i+\hat d}\cdot \hat d_2)],
\end{eqnarray}
and two at third neighbor
\begin{equation}
\label{eqn39}
+{1\over N}\sum_i\sum_{\hat d=nn3}[
f_L(\vec\xi_i\cdot \hat d)(\vec\xi_{i+\hat d}\cdot \hat d)
+f_T(\vec\xi_i\cdot\vec\xi_{i+\hat d}
-(\vec\xi_i\cdot \hat d)(\vec\xi_{i+\hat d}\cdot \hat d))].
\end{equation}
In these expressions, $\hat d$ is a unit vector that is summed
over all directions from a site $i$ to the neighbors in a given shell.
For second neighbors, a further distinction is made between
the two direction orthogonal to $\hat d$: $\hat d_2$ is the
Cartesian vector orthogonal to $\hat d$, and $\hat d_1$ is
orthogonal to both $\hat d$ and $\hat d_2$.
$\vec\xi_{i+\hat d}$ is the value of $\vec\xi$ for the site at distance 
$\vec d$
from site $i$.  

	By substituting the results of Table {\ref{harm.tbl}} 
and the appropriate
values of $\vec\xi_i$ into Eqs.~{\ref{eqn37}}-{\ref{eqn39}}, we obtained
the local harmonic coefficient $A$ and the intersite coupling constants
$b_L$, $b_T$, $d_L + d_{T1}$, $d_{T2}$ and
$f_{L} + 2 f_{T}$ shown in Table {\ref{param.tbl}}. 
There was not enough information at the high symmetry 
$\bf {q}$ points used to separate $d_L$ from
$d_{T1}$ or $f_{L}$ from $f_{T}$.
As long as $d_L$, $d_{T1}$, $f_{L}$ and 
$f_{T}$ were all nonnegative, however, we found that the values
chosen had little effect on the thermodynamic properties of
our model, so we set 
$f_{L} = f_{T}$ and $d_L = d_{T1}$. 
Note that the local correction to the dipole-dipole
energy decreases as the intersite distance
increases, becoming smaller than the dipole-dipole
interaction at third neighbors.  While there
is a tendency for the intersite interaction
to approach the dipole-dipole form at large distances,
there is still significant deviation from this
form at third neighbor distance.

	Next, consider the anharmonic terms to be included in 
the model Hamiltonian.  
Such terms
are crucial for determining the thermodynamics of a system undergoing a  
mode-softening structural phase transition.
Making an important simplifying approximation which has been previously 
applied in the study of GeTe\cite{Rab87} and perovskite 
oxides\cite{Rab95pt,WaghRab,Zho94bt}, we ignore multisite  
and higher-order pair interactions in our 
model Hamiltonian and include anharmonic terms only in the local 
distortion energy
$U_{la}$.
The Ge sites are centers of full cubic symmetry; thus
the onsite energy expansion includes only cubic invariant terms.
We truncate the expansion at eighth order, including all terms up to 
quartic order and the isotropic sixth and eighth order terms:
\begin{equation}
\label{eqn310}
U_{la}/N = {1\over N}\sum_i(B|\vec \xi_i |^4+
C(\xi_{ix}^4+\xi_{iy}^4+\xi_{iz}^4)
+D|\vec \xi_i |^6+E|\vec \xi_i |^8).
\end{equation}
The coefficients B, C, D and E were obtained from a series of frozen 
phonon calculations at
$\Gamma$,
after subtracting the 
$\Gamma$ point harmonic energy,
$-1.6524\, {\rm eV}\, |\vec\xi|^2$ (Table {\ref{harm.tbl}}).
Various amplitudes
of distortion
$|\vec\xi|$ were applied in both the 
$\hat{z}$ and
$(\hat{x} + \hat{y} + \hat{z})/\sqrt{3}$
directions and a least squares fit to Eq. {\ref{eqn310}} was
performed using the {\it ab initio} total energy results.
The resulting coefficients are given
in Table {\ref{param.tbl}}.
A comparison of the frozen phonon and fitted results is
shown in Figure {\ref{gammaen.ps}}(a).
A contour plot of the equal
energy surfaces of the fit is shown in Figure {\ref{gammaen.ps}}(b).
The energy minima are located along the [111] directions.

	While the spin-like model obtained so far includes the
necessary ingredients for a phase transition, the ground
state symmetry of the crystal, degree of distortion and
the nature of the phase transition all depend on
magnitudes of both the strain (Eq. {\ref{eqn23b}}) and 
strain coupling terms (Eq. {\ref{eqn23c}}).\cite{Wag97sc}
We include the effects of strain and strain coupling
to lowest order.  The symmetry adapted forms
are:
\begin{equation}
\label{eqn311}
U_{strain}/N =
{1\over 2}C_{11}\sum_\alpha e_{\alpha \alpha}^2
+{1\over 2}C_{12}\sum_{\alpha \neq \beta}e_{\alpha \alpha}e_{\beta \beta}
+{1\over 4}C_{44}\sum_{\alpha \neq \beta}e_{\alpha \beta}^2
\end{equation}
and
\begin{eqnarray}
U_{strain~coupling}/N =
\label{eqn312}
{g_0\over N}(\sum_\alpha e_{\alpha \alpha})\sum_i|\vec \xi_i |^2
& + & {g_1\over N}\sum_\alpha(e_{\alpha \alpha}\sum_i\xi_{i \alpha}^2)
\nonumber \\
 & + & {g_2\over N}\sum_{\alpha < \beta}e_{\alpha \beta}\sum_i\xi_{i  
\alpha}\xi_{i \beta}
\end{eqnarray}

	The variations in cell energy of uniformly 
strained Pb$_3$GeTe$_4$ structures, with Pulay corrections, are
shown in Figure \ref{strain.ps}
for the following types of strain:
uniform dilation ($e_{xx}$ = $e_{yy}$ = $e_{zz}$), tetragonal
strain ($e_{xx}$ = $e_{yy}$ = -- $e_{zz}$/2) and  rhombohedral shear
($e_{xy}$ = $e_{yx}$ = $e_{xz}$ = $e_{zx}$ = $e_{yz}$ = $e_{zy}$).
Least squares fits to these results determine the elastic
constants 
$C_{11}$, $C_{12}$ and $C_{44}$ shown in Table {\ref{param.tbl}}.

	Finally, strain couplings were calculated by 
subtracting the energy
at $\vec\xi = 0$ from various uniform fixed amplitude 
($|\vec\xi| = 0.25$) distortions
under various strains,
using Eq. {\ref{eqn312}} to determine the form of the corresponding 
energy differences
and finding the unknown coefficients via the linear parts of 
least-squares quadratic fits to each curve. 
The results are shown in Fig. \ref{couple.ps}.  
In Table {\ref{param.tbl}}, we present the 
values obtained for the three independent strain coupling parameters
$g_0$, $g_1$ and $g_2$. 
As a check, further calculations were performed in which the amplitude 
of the LWF distortion was also varied.  The results were
consistent with the strain coupling parameters found.

     The effect of strain on the energy of a $\Gamma$ local polar
distortion is shown in Figure \ref{gammacup.ps}, which is similar
to Figure \ref{gammaen.ps}(b), except that the distortion energy
is now minimized with respect to uniform strain.  There are still
8 local minima along the [111] directions.  Their energies 
are 17 meV/cell lower, and corresponding values of 
$|\vec\xi_i|$ 0.035 larger than if the terms involving
strain are neglected.

\section{Finite Temperature Simulations}

	The model Hamiltonian constructed in the previous 
section is completely  
specified by
\begin{eqnarray}
\label{eqn41}
 U(\{ \vec\xi_i \},{\bf e})/N = (U_0 + U_{lh} + U_{dd} + U_{sr} +
U_{la} + U_{strain} + U_{strain~coupling})/N,
\end{eqnarray}
where the individual contributions are given by 
Eqs.~{\ref{eqn31}} and {\ref{eqn36}}-{\ref{eqn312}}
and the parameters appear in Table {\ref{param.tbl}}.
This model Hamiltonian applies to all ionic displacements
believed to play a role in the structural phase transition
and allows one to calculate T$_c$ and other properties in the
vicinity of the transition for
the ordered Pb$_3$GeTe$_4$ system.  
We analyzed the finite-temperature behavior using a classical ``single
flip" Metropolis Monte Carlo 
simulation.   
Three types of Monte Carlo steps were applied:
(1) A ``spin" $\vec\xi_i$ was chosen at random.  A random vector 
located within a cube centered on the origin was added to the chosen 
spin vector.  Acceptance ratios of about $1/e$ were obtained
for a cube radius of $2 \sqrt{kT}$, where $kT$ is in eV.
(2) A spin was chosen at random.  The $x$, $y$ and $z$ components
were independently multiplied by $\pm 1$ with probability 0.5.
This led to a great increase in simulation speed as 
the spins were able to overcome barriers between local energy
minima. 
An acceptance ratio of about 0.3 was found
for this step near $T_c$.
(3) One strain component was chosen at random.  A random 
number between $-\epsilon$ and $+\epsilon$ was added to this strain
component.  Acceptance ratios of about $1/e$ were obtained for
cubic cells for
$\epsilon = 0.15 \sqrt{kT/L}$, where $kT$ was in eV and $L$
was the size of the cube edge in units of $a$.

	We simulated an $L \times L \times L$ supercell of
cubic Pb$_3$GeTe$_4$, with $L = 10$, representing 8000 atoms.  
We determined mean total energies and lattice
parameters as a function of temperature. 
At each temperature, the parameters were averaged over
$N_s$ Monte Carlo attempts for $N_c = 100$ cycles.  
The longest autocorrelation
time for the ${\bf q} = (2 \pi/L) \hat{z}$  Fourier component
of the spin field was determined to estimate the
decorrelation time $\tau$ at each temperature,
For a proper ensemble
average, $N_s N_c >> \tau$.  We generally selected
$N_s$ so that $N_s N_c \approx 20 \tau$.
From the root mean
square deviation of the individual cycle averages, 
$\sigma_{rms}$, we can obtain an estimate of the
$rms$ error for the ensemble averaged quantities\cite{Bin76},
\begin{equation}
\sigma = \sqrt{1 + (\tau/N_s)^2} (\sigma_{rms})/(\sqrt{N_c - 1}).
\end{equation}
We ran the simulations for a
range of temperatures between 0K and 1000K.  We began at 1000K, 
cooled to 0K and heated back up to test for hysteresis effects.  
At each temperature, the system was allowed to equilibrate before 
any sampling began.  

  At each temperature, the symmetry of the average cell was either
cubic or rhombohedral to within statistical errors.  Therefore
the lattice parameter and the rhombohedral angle as a function
of temperature are sufficient to give the unit cell as a function
of temperature.
The results are shown in 
Figure \ref{monte.ps} and are consistent with a second order 
cubic-rhombohedral
phase transition
at T$_c$ = 620K. 
The error bars for certain
temperatures very near $T_c$ may be consistent with slight
hysteresis, indicating a very weak first order transition.  However,
since this is observed in the temperature range where the
effects of the periodic boundary conditions are significant, no
firm conclusion can be reached.  In any case, the
hysteresis effect, if it exists, spans less than 10 degrees.
Experiments on disordered Pb$_{1-x}$Ge$_x$Te
for small $x$ show a continuous transition\cite{Hoh72}
and give $T_c \approx 350K$ for $x = 0.25$.
This apparent discrepancy in $T_c$ will be discussed in detail
in the next section.

  The plot of lattice parameter versus temperature shows a 
discontinuous slope at $T_c$ and a negative coefficient of
thermal expansion below $T_c$.  The model cannot be expected
to obtain the right value for thermal expansion 
coefficient because it
neglects anharmonic terms involving the normal modes
complementary to the effective Hamiltonian subspace.
The inclusion of additional modes should not alter the predicted
{\it discontinuity} in the expansion coefficient,
however,
and a discontinuity of the right magnitude has 
indeed been observed experimentally 
in Pb$_{1-x}$Ge$_x$Te.\cite{Hoh72}

\section{Discussion}

	  In this section, we discuss further various 
features of our model Hamiltonian, compare the model with 
other systems and models, comment on the apparent
discrepancy between our $T_c$ and the experimental value, 
and discuss what the present calculations have taught 
us about generalizing to the case of {\it disordered}
systems.

  The harmonic part of our model is completely 
given by the normal mode dispersion curves.
In Figure
{\ref{dispersion.ps}}, these dispersion curves are shown for
high symmetry paths in the Brillouin zone.  Three features
are particularly noteworthy: (1) Roughly speaking, there is
a relatively flat branch at $\omega \approx 425$ cm$^{-1}$ and 
a relatively flat branch at $\omega \approx 600 i$ cm$^{-1}$.
This reflects the dominance of the longitudinal first
neighbor dipole-dipole interaction and correction term
$b_L$.  (2)  Although weaker than the first neighbor 
longitudinal interaction, the other interactions are
crucial for determining the ground state structure.  For
example, if the correction to the first neighbor transverse
interaction, $b_T$, were zero, the most unstable modes
would have symmetry $M_{1^{\prime}}$ and the ground state would
be antiferroelectric rather than ferroelectric.
(3) The long-range dipole-dipole interaction leads to LO-TO
splitting at $\Gamma$.  The frequency of the TO mode at
$\Gamma$ is still negative, as opposed to the case of 
various perovskite ferroelectrics\cite{Zho94}, where the LO
mode corresponding to the soft TO mode generally has very
high frequency.  This is a consequence of the strength of 
the Ge off-centering instability in Pb$_{1-x}$Ge$_x$Te.

  The form of our model Hamiltonian is that of interacting
vectors on a cubic lattice.  The same form of model
Hamiltonian applies to perovskite ferroelectrics; thus
the difference in phase transformation behavior of the
different models can be related to the differences in
the parameters and/or the number of vectors per unit
cell.  A comparative study of all the models to date is
outside the scope of this paper; we will just compare the
effects of strain coupling in our Pb$_3$GeTe$_4$ model and
the PbTiO$_3$ model of Waghmare and Rabe.\cite{Rab95pt}
The effect of strain coupling in Pb$_3$GeTe$_4$ is much weaker than
in PbTiO$_3$.  For each case, we compared the energy in the 
ground state of the corresponding model with the symmetric
reference state energy.  In Pb$_3$GeTe$_4$, the
distortion energy was --121 meV/cell, while strain and
strain coupling contributed only --20 meV/cell.
In PbTiO$_3$ on the other hand, the strain and strain coupling 
energy was --233 meV/cell at the ground state.  Neglecting these
terms, the distorted ground
state was 120 meV/cell {\it higher} in energy than the 
reference state.  
The difference in the strength of strain coupling also makes the
difference between a continuous phase transition in our
Pb$_3$GeTe$_4$ model and a first-order phase transition in
the PbTiO$_3$ model.\cite{Rab95pt,Wag97sc}

  In the Ge off-centering model of Pb$_{1-x}$Ge$_x$Te\cite{Log77,Kat80},
each Ge atom is displaced off-center in one of the 
8 cubic [111] directions.  The cubic-rhombohedral
phase transition is an order-disorder transition involving
the {\it directions} of the displacements.
Above $T_c$, the directions of the Ge atom
displacements is disordered.  There is no net polarization and 
the symmetry is cubic. 
Below $T_c$, the directions of the Ge atom
displacements are ordered and lie predominantly
along one axis, lowering the symmetry from cubic
to rhombohedral and making the system polar.

    Our model, derived from first principles, 
essentially confirms the Ge off-centering model.
It contains a local polar distortion (the LWF) 
dominated by Ge motion, whose energy minima are in directions 
where the Ge are displaced along the [111] directions 
(Figure {\ref{gammaen.ps}}) in agreement with the off-centering model. 
To see if the phase transition in our model is order-disorder, 
we calculated the
distribution of local distortion vectors at $T_c$ -- 50K
and $T_c$ + 50K.
The results are shown in Figure {\ref{distrib.ps}}.
Above $T_c$, there are distinct peaks in all
eight [111] directions, while below $T_c$, there is only
a single peak in one of the [111] directions, confirming
the order-disorder nature of the transformation.
We note here that EXAFS measurements of 
Islam and Bunker\cite{Isl87} leave an ambiguity in the 
direction of displacements of
the Ge atoms at low temperature.
For a ground state with rhombohedral strain along
the (111) direction, say, both (111) and $(\overline{111})$ local
polar distortions lead to the same Ge-Te near neighbor
pair distribution functions.  In our model, the
interactions between the local polar distortions
have been determined and they are found 
to lead unequivocally to strictly ferroelectric
ordering of these distortions.

EXAFS
measurements\cite{Isl87} show two peaks in the Ge-Te
distribution function in Pb$_{1-x}$Ge$_x$Te.  Only a displacement of
Ge along one of the [111] directions breaks the symmetry
in a way that leads to two different Ge-Te
near neighbor distances.  Although the local polar
distortion of our model involves motion of atoms
other than the central Ge, the same result holds
that only a local polar distortion along a [111]
direction leads to exactly two different Ge-Te
distances.  However, in our model, both the
fact that the
ionic displacement patterns corresponding to
local polar distortion on different Ge atoms
can overlap and the fact that the
local polar distortions do not lie strictly along
the
[111] directions (see Figure {\ref{distrib.ps}})
makes it less certain that there is a two-peak
Ge-Te distribution function.
By transforming the
$\{ \vec\xi_i \}$ coordinates of
our model back to ionic displacements via equation \ref{eqn28}, we can
calculate the first-neighbor Ge-Te distribution functions
of our model.  We show our results in Figure {\ref{distrib.ps}}.
There are two peaks, even above $T_c$, although
there is significant overlap. These distribution functions
do not incorporate the thermal noise of the neglected
normal modes, which would tend to further broaden the peaks;
nonetheless, they are consistent with the experimental results
showing two peaks.

  Since the semi-empirical models of Katayama and
Murase\cite{Kat80} (KM) and of Yaraneri {\it et al.}\cite{Yar81}
are based upon Ge ions tunneling between (111) and $(\overline{111})$
positions, in apparent contradiction with our results 
showing one preferred position below $T_c$, 
we have investigated this issue more carefully.
Extrapolating our model to lower Ge concentrations by setting
the dipole-dipole interactions to be that corresponding to
an fcc dipole lattice at the given density and interpolating 
the lattice parameter between the LDA values for PbTe and
Pb$_3$GeTe$_4$ via Vegard's law, we obtain a rough estimate of
composition $x \approx 0.05$, below which there are two
wells in the potential along the (111) line.  
Using a different analysis, Katayama and Murase
estimated the range of validity of their model
to be $x < 0.1$.

	Although our model is classical, it sheds some light on
quantum tunneling phenomena in the Pb$_{1-x}$Ge$_x$Te system at small $x$ 
by providing a potential energy surface for Ge motion.  
In Figure \ref{tunnel.ps}(a), we show the potential energy surface for 
Ge motion in cubic Pb$_3$GeTe$_4$ at zero temperature. There is
only a single deep well for the Ge ion and thus no quantum
tunneling.
A more relevant potential energy surface is that for
a single Ge impurity in PbTe.   While our calculations were
performed on a structure with $25\%$ Ge, a crude extrapolation
to the isolated impurity case is possible by (1) setting all
Ge-Ge interaction terms equal to zero and (2) setting the
lattice parameter to the value for pure PbTe.  When this
is done, the potential energy surface for Ge atom motion is
that shown in Figure \ref{tunnel.ps}(b).  There are now eight wells 
with minima along cubic [111] directions, about 7 meV deep with
respect to the centered position and with 1 meV barriers between wells.
The zero-point vibrational energy of a Ge atom in this potential
is about 20 meV, so it would not be localized in a single well.
At some nonzero concentration of Ge impurities, we expect the
Ge-Ge off-centering interactions to overcome the zero-point
vibrational motion leading to an ordered rhombohedral ground state.

  The KM model includes not only
Ge off-centering, but coupling of Ge motion to the PbTe
transverse optical phonons.
In our calculation of normal modes,
the coupling of Ge motion to other ionic motions is
given in the dynamical matrix, which is then diagonalized.
Compared to the KM model,
we both reduce the number of degrees of freedom
and incorporate normal modes
that are more appropriate for a material containing
substitution.  Consider our $\Gamma_{15}$ eigenmode 
(Table \ref{mode.tbl}).  All Pb motion is this eigenmode is 
in the opposite direction of all Te motion,
exactly as would be the case of the zone center PbTe TO
phonon.  The motion of one Te atom, however, is
much larger in magnitude than all other Pb and Te
motions, reflecting the importance of localization
phenomena. 

	It is instructive to compare our results for Pb$_3$GeTe$_4$ with
similar results for pure PbTe.  Linear response calculations
on PbTe were carried out using the experimental rocksalt structure 
as the reference structure and the LDA lattice parameter.
All normal modes are stable,
consistent with the experimental fact that PbTe is
cubic down to zero temperature.\cite{Hoh72,Tak79}  It is also 
consistent with the fact that the instabilities in ternary Pb$_3$GeTe$_4$ are
associated with the Ge ions.

 	For pure PbTe, we obtain Born effective charges of +5.84
for Pb and --5.84 for Te and an electronic dielectric constant of 33.0.
These are in excellent agreement with the experimental values
of $Z^{\star} = \pm 6.0$ and $\epsilon_{\infty} = 32.8$\cite{Bur70}
and suggest that our values for Pb$_3$GeTe$_4$ are also reliable.
Our calculations show that $\epsilon_{\infty}$ for Pb$_3$GeTe$_4$ is
about 40\% higher than for PbTe.  
When Ge is substituted for Pb, the Pb effective charge changes
little and remains nearly isotropic. 
The Ge effective charge itself is significantly larger 
than that for the Pb it replaces.  
The Te(2) Z$^{\star}$ also changes substantially and becomes
markedly anisotropic.  An 
analogous anisotropy has been observed for the oxygen Z$^{\star}$ in ABO$_3$
perovskite ferroelectrics.\cite{Zho94,Gho95}
It is interesting that the two Z$^{\star}$ components of highest
magnitude are associated with the two most significant ionic
displacements in the LWF, namely Ge off-centering along with axial
motion of the first neighbor Te ions.

  	The values of the Born effective charges (and
$\epsilon_{\infty}$) in Pb$_3$GeTe$_4$ do not remain constant as the
structure distorts from its reference structure to its ground
state.  We set the structure of Pb$_3$GeTe$_4$ to the ground state of
our model ($e_{\alpha\alpha} = 6.37 \times 10^{-3};
e_{\alpha\beta} = 6.03 \times 10^{-3}, \alpha \neq \beta;
\vec\xi = (0.255,0.255,0.255))$ and recalculated the Born
effective charge tensors.  The results are shown in Table
\ref{bornagain.tbl}.  The magnitude of the Born effective charges are
generally smaller in the distorted state, sometimes much smaller.
In particular, $Z^{\star}$ for Ge along (111) goes from +8.02 to
+1.52; the {\it electronic} contribution to Z$^{\star}$ goes from positive to 
negative.  Previous studies have shown similar, but smaller decreases in
magnitudes of $Z^{\star}$ in BaTiO$_3$\cite{Gho95} and
KNbO$_3$\cite{Wan96} as the structure is distorted.
A common trend is that the atoms with the largest relative motion in
the structural instability have the largest changes in $Z^{\star}$.
The magnitude of 
$\epsilon_{\infty}$
also decreased substantially as the structure distorted.  A similar,
but relatively smaller,
decrease has been calculated for KNbO$_3$\cite{Wan96}.

	The large difference between $T_c$ for our Pb$_3$GeTe$_4$ model and the 
experimental $T_c$ for the disordered system at the same 
composition merits discussion.  The first source for
the discrepancy is the approximations used in the 
first-principles
energy calculations.
Approximations in the ab initio calculations include the use
of pseudopotentials, LDA, finite k-point sampling, a finite
cutoff energy and a finite real-space grid for charge density.
The second source for the
discrepancy is the net effect of the neglected anharmonic
effects, which may be important in Pb$_3$GeTe$_4$
because the relative displacements of the ions is so
large (order 0.5~\AA).
For example, the dependence of $Z^{\star}$ on distortion
discussed in the previous paragraph
will lead to a nonlinear dependence of $\bf{P}_{i}$ on
$\vec\xi_i$, which in turn will lead to anharmonic intersite
interactions.
Such anharmonic effects could either raise or
lower $T_c$ and are worth further study.
The third source for the discrepancy is our
use of the LDA lattice parameter rather the (unknown) experimental
one.\cite{Yin82}.  
Typically,
{\it ab initio} phonon calculations using
experimental lattice parameters are more accurate than
those based on the LDA lattice parameters.\cite{Yin82}
In section 3, we estimated that the LDA 
lattice parameter $a$ for Pb$_3$GeTe$_4$ is about 0.5\% smaller than
the experimental value, so more accurate results might be
obtained by raising $a$.
In general, raising the 
lattice parameter will ``soften" the phonons and thus raise $T_c$;
in the present case, this would increase the 
discrepancy between our $T_c$ and the experimental value.

	Finally, there is the problem of extending the results for
an ordered structure to those for a disordered alloy.
We have, in effect, considered only one term in Eq. {\ref{eqn21}},
which is not in fact ``typical".
For example, in
the cubic configuration 
no relaxation is possible.   Other configurations at the same
composition do allow relaxation.
We have preliminary {\it ab initio} results that show that the 
relaxation in disordered Pb$_{1-x}$Ge$_x$Te is dominated by a 
0.11 to 0.14~\AA~inward displacement of each Te in the first 
neighbor shell of each Ge ion, where such relaxation does not
conflict with the relaxation about a second Ge ion.
Such relaxation alone lowers the energy per
Pb$_3$GeTe$_4$ unit as much as 20 meV per atom with respect to 
the undistorted cubic structure.
Locally, the relaxation mimics a decrease of the lattice
parameter, which should ``harden" the unstable Ge-dominated
modes and thus decrease $T_c$.

	The large energy differences between different configurations
of Pb$_3$GeTe$_4$ upon relaxation shows the necessity of including 
different chemical
configurations in order to correctly model the structural
phase transition in the disordered system. 
This is similar to the case of
alloy phase
diagrams\cite{Con83}, where order-disorder transitions
predominate.  There, it has been shown that lattice entropy
effects must be included to obtain correct transition
temperatures for order-disorder transitions.\cite{Sil95,Gar96}
Here, we need to include configurational entropy in order
to study a structural phase transition.

  The situation for Pb$_{1-x}$Ge$_x$Te is complicated by
the existence of a miscibility gap, with the peak of
the exsolution dome at about 840 K\cite{Hoh72}.
For $x = 0.25$, as in this work, phase separation
to Ge-rich and Pb-rich phases occurs at about 770 K\cite{Hoh72},
although a mixed phase of this composition that is 
metastable at low temperatures can be obtained by rapid
quenching\cite{Hoh72}.  Thus, the experimental result of
$T_c \approx 350 K$ for the structural phase
transition in Pb$_{0.75}$Ge$_{0.25}$Te actually applies to a 
metastable compound and the thermodynamic theory for the
disordered compound is complicated by the fact that the
equilibrium partition function (Eq. {\ref{eqn21}}) does
not apply.

	Finally, we discuss the form of the model that
should describe disordered Pb$_{1-x}$Ge$_x$Te.  
For each configuration in Eq.~\ref{eqn21}, it is possible
in principle to repeat the procedure used in this work
to develop a configuration specific model.
Based on the Ge off-centering picture, we make the following
conjectures about the nature of all such models. (1) 
There is a basis for all unstable modes that involves a vector LWF centered
on each Ge. (2) The displacement pattern corresponding to
each LWF will be strongly localized. (3) The interactions between
local polar distortions centered on different Ge will favor ferroelectric
ordering.

	Given enough models for individual configurations, it
should be possible to extract a ``supermodel" that applies
to all Pb$_{1-x}$Ge$_x$Te configurations.  In the same way
that energy for a disordered alloy can be given in terms of a 
cluster expansion\cite{Con83}, the values of the model parameters could
also be given in terms of cluster expansions.
Work is in progress to develop models for other Pb$_{1-x}$Ge$_x$Te
configurations and to incorporate the effects of relaxation into
these models.

\section{Conclusions}

	As a first step towards a first-principles study of the effect 
of substitutional disorder in
Pb$_{1-x}$Ge$_x$Te, we have developed a model Hamiltonian for an
ordered Pb$_3$GeTe$_4$ system.  We have determined the parameters in this
model from {\it ab initio} calculations.  From a classical Monte Carlo
simulation, $T_c$ was found to be about 620K.
Comparison of the present results with those obtained for other
configurations will demonstrate how transferable the
parameters of the model Hamiltonian for this  
ordered supercell are to other configurations, show
how much $T_c$ depends on configuration.
and 
allow a model for the disordered system to be developed.

\section{Appendix}

	The use of a plane wave basis set and periodic boundary
conditions in an {\it ab initio} calculation
introduces two kinds of error. The first is due to the
finite energy cutoff for the plane waves; the second is due to the
discretization of wavevectors that is a result of using a finite
periodic cell in real space.  In this paper, we use Pulay
corrections to compensate for the second source of error.

	The specific method for applying Pulay corrections used here was
taken from Rignanese {\it et al.}\cite{Rig95}.  In this method,
the density functional total energy $E_{tot}$ of a periodic system at
fixed volume $V_0$ is measured at several cutoff energies
$E_{cut}$ and treated as a function of the average number of
plane waves per {\bf k} point, $\overline N_{PW}$
at those values of $E_{cut}$.
Rignanese {\it et al.} suggest using cutoff energies of
$E_{cut}$ (the fixed value for a set of calculations of different
cells), $E_{cut} - 3\% $ and $E_{cut} + 3\% $ and then fitting
through the resulting total energies
$E_{tot}[\overline N_{PW}, V_0]$
via the following function:
\begin{equation}
E_{tot}[\overline N_{PW}, V_0] = E^{inf}_{tot} + {\rm exp} 
(a_0 + a_1 \overline N_{PW}).
\end{equation}
Finally, the results obtained for the {\it single} cell volume $V_0$ 
are used
to determine the Pulay corrections to total energy for {\it any} cell
volume $V_1$ via the correction:
\begin{equation}
E^c_{tot} [E_{cut},V_1] \sim E^d_{tot} [E_{cut},V_1] +
E_{tot} [{V_0\over{V_1}} \overline N^c_{PW} (E_{cut}, V_1), V_0] -
E_{tot} [{V_0\over{V_1}} \overline N^d_{PW} (E_{cut}, V_1), V_0]
\end{equation}
$E^d_{tot} [E_{cut},V_1]$ is the total energy measured without 
corrections.
$\overline N^d_{PW}$ is the corresponding average number of plane
waves per $\bf k$ point.  
$\overline N^c_{PW}$ is the ``continuous" number of plane waves
at $E_{cut}$, determined by multiplying the density of states in
reciprocal space by the volume of the sphere containing plane waves
of energy less than $E_{cut}$ and is given by 
\begin{equation}
\overline{N}^c_{PW} (E_{cut},V) = {V\over{6 \pi^2}} (2 E_{cut})^{3/2},
\end{equation}
when $V$ and $E_{cut}$ are in atomic units.
From the value of $\overline{N}^c_{PW} (E_{cut},V_1)$ obtained, the
corrected total energy $E^c_{tot} [E_{cut},V_1]$ can be computed.

  By measuring the total energy of Pb$_3$GeTe$_4$
at $a = 6.375$~\AA~and PbTe at $a = 6.275$~\AA~at~$E_{cut}$ = 291, 300 
and 309 eV, the following
fitting forms were obtained:
\begin{eqnarray}
\label{eqn74}
{\rm PbTe}: E_{tot} [\overline N_{PW}, V_0 = 259.05 \AA^3] 
& = & -1296.5850 + 
6.244~{\rm exp}(-1.956 10^{-4} \overline N_{PW}^d) \\
{\rm Pb}_3{\rm GeTe}_4:  E_{tot} [\overline N_{PW}, V_0 = 247.05 \AA^3] 
& = & -1308.4464 + 
8.452~{\rm exp}(-9.307 10^{-5} \overline N_{PW}^d)
\nonumber
\end{eqnarray}
These results were used to correct
all the CASTEP~2.1 total energy results used in this article.
From the expression ({\ref{eqn74}}) for $E_{tot}$, and the
value of $\overline N_{PW}^d = 2911.75$ corresponding 
to our calculation
for Pb$_3$GeTe$_4$ at $a = 6.275$~\AA, we obtained
$E_{tot} \equiv U_0/N =  -1302.1725~{\rm eV}$.

\section{ Acknowledgements}

  This work was supported by ONR N00014-91-J-1247.
We thank Philippe Ghosez and Umesh Waghmare 
for useful discussions and helpful assistance. 
We thank Alex Elliott and Nicola Hill for their comments on the
manuscript.

\newpage
\baselineskip = 2\baselineskip  

\begin{figure}
\caption{Crystal structures of the observed phases of pure GeTe. Eight 
atom  regions are shown for (a) the high-temperature rocksalt
structure and (b) the low-temperature rhombohedral structure, which is 
related to (a) by a rhombohedral lattice distortion and a 
relative displacement of the sublattices, shown by arrows. Both 
distortions are greatly exaggerated for clarity.}
\label{gete.ps}
\end{figure}
 
\begin{figure}
\caption{The high-symmetry reference structure of the ordered cubic 
Pb$_3$GeTe$_4$  eight-atom unit cell, produced by replacing a cubic 
superlattice of Pb ions (striped circles) in the PbTe rocksalt structure 
with Ge ions (solid circles).}
\label{structure.ps}
\end{figure}

\begin{figure}
\caption{The $z$ component of the estimated lattice Wannier
function of cubic Pb$_3$GeTe$_4$ (doubled for clarity). Only ions
with displacements greater than 0.1 times
the displacement of the central Ge ion are shown.
Note that
the unit cell outlined is translated with respect to
that in 
Figure 2.
}
\label{lwf.ps}
\end{figure}

\begin{figure}
\caption{(a) Comparison of frozen phonon and fitted results
for the unstable $\Gamma$ mode subspace energy; strain fixed
at zero.
(b) Contour plot of $1\overline{1}0$
plane cross section of eighth order expansion of unstable $\Gamma$ 
mode subspace energy.  The contour interval is 20 meV and the grid
spacing is 0.2.}
\label{gammaen.ps}
\end{figure}

\begin{figure}
\caption{Energy vs. strain for (a) uniform dilation ($e_{xx} = e_{yy} = 
e_{zz}$), (b) tetragonal strain ($e_{xx} = e_{yy} = -e_{zz}/2$) and 
(c) rhombohedral strain ($e_{xy} = e_{xz} = e_{yz}$).}
\label{strain.ps}
\end{figure}

\begin{figure}
\caption{Energy change due to local polar distortion of amplitude 0.25
on strained cells.  (a) uniform dilation; $\vec\xi = |\vec\xi| \hat{z}$,
(b) tetragonal strain; $\vec\xi = |\vec\xi| \hat{z}$,
(c) rhombohedral strain; $\vec\xi = |\vec\xi| (\hat{x} +
\hat{y} + \hat{z})/\sqrt{3}$.}
\label{couple.ps}
\end{figure}

\begin{figure}
\caption{Contour plot of $1\overline{1}0$ plane cross section of 
unstable $\Gamma$ mode subspace energy, minimized with respect to 
uniform strain. 
The contour interval and grid spacing are the same as in 
Figure 4.}
\label{gammacup.ps}
\end{figure}

\begin{figure}
\caption{Result of Monte Carlo simulations. (a) Rhombohedral strain and 
(b) lattice parameter versus temperature.
The results are consistent with a second order
phase transition with $T_c$ near 620 K.}
\label{monte.ps}
\end{figure}

\begin{figure}
\caption{Phonon dispersion relations for the cubic Pb$_3$GeTe$_4$ model
Hamiltonian.}
\label{dispersion.ps}
\end{figure}

\begin{figure}
\caption{Distribution of $\vec\xi$ in Monte Carlo simulation. 
(a) Contour plot in
$1\overline{1}0$ plane at $T_c$+50K.  (b) Same as (a), at T = $T_c$--50K.
(c) First neighbor Ge-Te distribution function at T = $T_c$+50K. (d) Same 
as (c) for T = $T_c$--50K.  Compare (a) with figure 4(b).  The eight [111] 
directions are preferred for local distortions, even above $T_c$, in 
agreement with the Ge off-centering picture.}
\label{distrib.ps}
\end{figure}

\begin{figure}
\caption{(a) Potential energy surface seen by single Ge ion at $T = 0$ 
in Pb$_3$GeTe$_4$ model (contour interval 50 meV).
(b) Potential energy surface seen by isolated Ge impurity in PbTe, based
on extrapolation of our model (contour interval 5 meV).}
\label{tunnel.ps}
\end{figure}

\newpage
\baselineskip = .5\baselineskip  
 
\begin{table}
\caption{Ab initio frequencies (in cm$^{-1}$) of selected normal modes in
cubic Pb$_3$GeTe$_4$.
The frequencies for optical $\Gamma_{15}$ modes are for the
transverse modes.}
\begin{tabular}{||l||c|c|c|c|c|c|c||}
\hline\hline 
Symmetry Label:  & $\Gamma_{15}$ & $\Gamma_{25}$ & $X_{1^{\prime}}$ &  
$X_{5^{\prime}}$ &
 $M_{1^{\prime}}$ & $M_{5^{\prime}}$ & $R_{15}$ \\
\hline & 648 i  & 176 & 478 & 593 i & 547 i & 395 & 429 \\
\hline &  0     & 630 & 518 & 163   &  315  & 603 & 437 \\
\hline &  195   &     & 721 & 308   &       & 688 & 683 \\
\hline &  292   &     &     & 622   &       & 736 &    \\
\hline &  518   &     &     &       &       &     &     \\
\hline &  662   &     &     &       &       &      &    \\
\hline\hline
\end{tabular}
\label{freq.tbl}
\end{table}

\begin{table}
\caption{Displacements (in \AA) associated with all symmetry-independent 
modes incorporated into the model Hamiltonian.
The periodicity of each
mode listed except $\Gamma_{15}$ is longer than the primitive cell and
involves the opposite motion of some ions in neighboring cells.}
\begin{tabular}{||l|c|c|c|c|c|c||}
Mode & $\Gamma_{15}$ & $X_{1^{\prime}}$ & 
$X_{5^{\prime}}$ & $M_{1^{\prime}}$ & $M_{5^{\prime}}$ & $R_{15}$ \\
\hline
$\omega$ (cm$^{-1}$) & 648 i & 478 & 593 i & 547 i & 395 & 429 \\
\hline
\bf{q} ($2 \pi/a$) & (0,0,0) & (0,0,0.5) & (0,0.5,0) &
(0.5,0.5,0) & (0,0.5,0.5) & (0.5,0.5,0.5) \\
$\bf u$ (Ge(0,0,0)) &  0.8878 $\hat{z}$ & 0.8258 $\hat{z}$ 
&  0.8835 $\hat{z}$ & 0.8654 $\hat{z}$ & 0.8207 $\hat{z}$ & 
0.8578 $\hat{z}$ \\
$\bf u$ (Te(0,0,0.5)) & $-0.4467 \hat{z}$ & 0 & $-0.4668 \hat{z}$ & 
$-0.5010 \hat{z}$
& 0  & 0 \\
$\bf u$ (Te(0,0.5,0)) & $-0.0575 \hat{z}$ & 0.0680 $\hat{z}$ & 0 & 
0      & 0      & 0 \\
$\bf u$ (Te(0.5,0,0)) & $-0.0575 \hat{z}$ & 0.0680 $\hat{z}$ 
& $-0.0318 \hat{z}$ & 0 & $0.0555 \hat{z}$      & 0 \\
$\bf u$ (Pb(0.5,0.5,0)) & 0.0156 $\hat{z}$ & $-0.5556 \hat{z}$ & 0 & 
0 & 0 & 0 \\
$\bf u$ (Pb(0.5,0,0.5)) & 0.0284 $\hat{z}$ & 0        & 0.0237 $\hat{z}$ 
& 0      & 0      & $-0.3569 \hat{x}$ \\
$\bf u$ (Pb(0,0.5,0.5) &  0.0284 $\hat{z}$ & 0      &      0 & 0     & 
$0.5684 \hat{y}$    & 
$-0.3569 \hat{y}$ \\
$\bf u$ (Te(0.5,0.5,0.5)) & $-0.0611 \hat{z}$ & 0   &  0     & 0      & 
$-0.0163 \hat{y}$     & 0 \\
\hline\hline
\end{tabular}
\label{mode.tbl}
\end{table}

\begin{table}
\caption{Ionic displacement pattern corresponding to z component of 
estimated lattice Wannier function $w_{z {\bf R}}$
for c-Pb$_3$GeTe$_4$.
The lattice Wannier function transforms according to the
$\Gamma_{15}$ representation for point group $O_h$
with center $R$.
Orbits of equivalent atoms for $w_{z {\bf R}}$ are 
labeled according to the
{\it International Tables}$^{56}$
convention for point 
group $4/mmm$.
Components whose values are 0 vanish by symmetry,
while those whose values are 0.0 are only zero by
approximation.}
\begin{tabular}{||c|c|c|c||}
Atom & ${\bf r}_i - {\bf R}$ (units of $a$)  & Multiplicity (orbit) & 
$\bf u$ (\AA) \\
\hline
Ge & (0,0,0)       & 1  &  $(0, 0, 0.8557)$ \\
Te & (0,0,0.5)     & 2(a)  & $(0, 0, -0.2352)$ \\
Ge & (0,0,1)       & 2(a)  & $(0, 0, 0.0122)$ \\
\hline
Te & (0.5,0,0)     & 4(c)  &  $(0, 0, 0.0043)$ \\
Pb & (0.5,0,0.5)   & 8(f)  &  $(0.1165 ,0, 0.0065)$ \\
Te & (0.5,0,1)     & 8(f)  & $(0.0, 0, -0.0133)$ \\
\hline
Pb & (0.5,0.5,0)   & 4(b)  &  $(0, 0, -0.0675)$ \\
Te & (0.5,0.5,0.5) & 8(e)  & $(-0.0020, -0.0020, -0.0076)$ \\
Pb & (0.5,0.5,1)   & 8(e)  & $(0.0,0.0, 0.0357)$ \\
\hline
Ge & (1,0,0)       & 4(c)  & $(0, 0, -0.0006)$ \\
Te & (1,0,0.5)     & 8(f)  &  $(0.0, 0, 0.0034)$ \\
Ge & (1,0,1)       & 8(f)  &  $(0.0, 0, 0.0017)$ \\
\hline
Te & (1,0.5,0)     & 8(d)  & $(0, 0, -0.0008)$ \\
Pb & (1,0.5,0.5)   & 16(g) &  $(0.0, 0.0128, 0.0003)$ \\
Te & (1,0.5,1)     & 16(g) & $(0.0, 0.0 -0.0012)$ \\
\hline
Ge & (1,1,0)       & 4(b)  & $(0, 0, 0.0009)$ \\
Te & (1,1,0.5)     & 8(e)  & $(0.0, 0.0, -0.0004)$ \\
Ge & (1,1,1)       & 8(e)  & $(0.0, 0.0, -0.0009)$ \\
\hline
\hline\hline
\end{tabular}
\label{lwf.tbl}
\end{table}

\begin{table}[h]
\caption{Born effective charges for 
cubic Pb$_3$GeTe$_4$.  For Ge and Te(1), $Z^{\star}$ is
isotropic, while for Pb and Te(2), the parallel 
and perpendicular components refer to
the local tetragonal axis.}
\begin{tabular}{||c|c|c||}
Species & $Z^{\star}_{\parallel}$ & $Z^{\star}_{\perp}$ \\
\hline
Pb & +6.15 & +6.06 \\
\hline
Ge & +8.02 & +8.02 \\
\hline
Te(1) & $-5.82$ & $-5.82$ \\
\hline 
Te(2) & $-7.73$ & $-6.37$ \\
\hline\hline
\end{tabular}
\label{born.tbl}
\end{table}

\begin{table}
\caption{Harmonic energies of effective Hamiltonian modes
$(|{\vec \xi}| = 1$ on each Ge site).  Units are eV per unit
cell.  Displacements of ions are given in 
Table II.
The energy for $\Gamma_{15}$ was determined via
a frozen phonon calculation, the others are determined
via nonzero $\bf{q}$ linear response.}
\begin{tabular} {||l|p{1.5cm}|p{1.5cm}|p{1.5cm}|p{10.5cm}||}
Mode & U$_{harm}$ & U$_{dipole}$ & U$_{sr}$ & Energy via 
Eqs.~3.7-3.9 \\
\hline
$\Gamma_{15}$    & --1.652 & --0.383 & --1.269  & A + b$_L$ + 
2 b$_T$ + 2 (d$_L$ + d$_{T1}$) + 2 d$_{T2}$ + 4 (f$_L$ + 2 f$_T$)/3  \\
\hline
$X_{1^{\prime}}$ & 1.220  & 0.885   & 0.335 & A -- b$_L$ 
+ 2 b$_T$ -- 2 (d$_L$ + d$_{T1}$) + 2 d$_{T2}$ -- 4 (f$_L$ + 2 f$_T$)/3 \\
\hline
$X_{5^{\prime}}$ & --1.387 & --0.443 & --0.944  & A + b$_L$
-- 2 d$_{T2}$ -- 4 (f$_L$ + 2 f$_T$)/3  \\
\hline
$M_{1^{\prime}}$ & --1.206  & --0.489   & --0.717  & A -- b$_L$ 
-- 2 d$_{T2}$ + 4 (f$_L$ + 2 f$_T$)/3  \\
\hline
$M_{5^{\prime}}$ & 0.844  & 0.245  &  0.599  & A + b$_L$ 
-- 2 b$_T$ -- 2 (d$_L$ + d$_{T1}$) + 2 d$_{T2}$ + 4 (f$_L$ + 2 f$_T$)/3  \\
\hline
$R_{15}$         & 0.927  & 0.000  & 0.927   & A -- b$_L$ 
-- 2 b$_T$ + 2 (d$_L$ + d$_{T1}$) + 2 d$_{T2}$ -- 4 (f$_L$ + 2 f$_T$)/3  \\
\end{tabular}
\label{harm.tbl}
\end{table}

\begin{table}
\caption{Parameters in the effective Hamiltonian for Pb$_3$GeTe$_4$
(units of eV per unit cell).}
\begin{tabular}{||c|c||c|c||c|c||}
\hline\hline $~~A~~$     &  --0.1765~  & $b_L\qquad$  & --0.7922~    &  
$C_{11}\qquad$   & 215.8~\\
\hline$B$     &  4.501  &   $b_T\qquad$   & --0.2148~  &
$C_{12}\qquad$   &  9.374~ \\
\hline$C$     &  6.165~  &   $d_L + d_{T1}$~ &   0.0766~  &
$C_{44}\qquad$   &  108.1~
\\  \hline$D$     &  --7.263~  &     &          &
$g_{0}\qquad$    & --3.50~
\\  \hline$E$     &  4.239  &   $d_{T2}\qquad$&  $-2.98 \times 10^{-2}$~  &
$g_{1}\qquad$    & --12.4~  \\
\hline Z$^{\star}$     & 12.10~  &   $f_{L} + 2 f_{T}$ &
$2.55\times 10^{-2}$~  &  $g_{2}\qquad$    & --9.99~  \\
\hline $\epsilon_{\infty}$        & 46.7~  &  $a$       & 6.275~\AA   &
              & \\
\hline\hline
\end{tabular}
\label{param.tbl}
\end{table}

\begin{table}[h]
\caption{Born effective charges and $\epsilon_{\infty}$ for
ground state of cubic Pb$_3$GeTe$_4$.  Atomic positions are
given to the nearest tenth.  The principal directions
are given by the subscripts.  They are exact by symmetry
for Ge, Te(1) and $\epsilon_{\infty}$, but 
only approximate for
Pb and for Te(2).  The corresponding values for the high
symmetry reference structure are given in 
Table IV
and shown here in parentheses.}
\begin{tabular}{||c|c|c|c||}
\hline
Pb (0.0,0.0,0.5) $Z^{\star}$ & +5.86$_{001}$ (6.15) & 
+5.84$_{1\overline{1}0}$ (6.06) & +4.98$_{110}$ (6.06) \\
\hline
Ge (0.0,0.0,0.0) $Z^{\star}$ & +1.52$_{111}$ (8.02) & 
+3.89$_{1\overline{1}0}$ (8.02) & +3.89$_{11\overline{2}}$ (8.02) \\
\hline
Te(1) (0.5,0.5,0.5) $Z^{\star}$ & --5.67$_{111}$ (--5.82) & 
--5.82$_{1\overline{1}0}$ (--5.82) & --5.82$_{11\overline{2}}$ (--5.82) \\
\hline
Te(2) (0.5,0.5,0.0) $Z^{\star}$ & --3.65$_{001}$ (--7.73) & 
--5.00$_{1\overline{1}0}$ (--6.37) & --5.42$_{110}$ (--6.37) \\
\hline
$\epsilon_{\infty}$  & 31.0$_{111}$ (46.7)  & 
30.9$_{1\overline{1}0}$ (46.7) & 30.9$_{11\overline{2}}$ (46.7)   \\
\hline\hline
\end{tabular}
\label{bornagain.tbl}
\end{table}

\enddocument
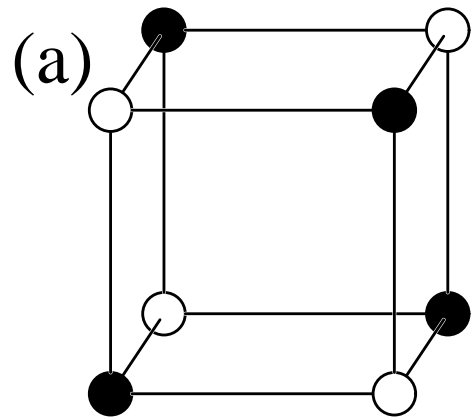
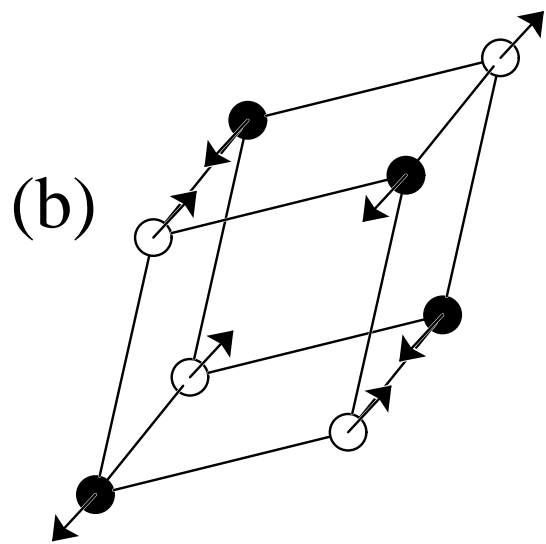
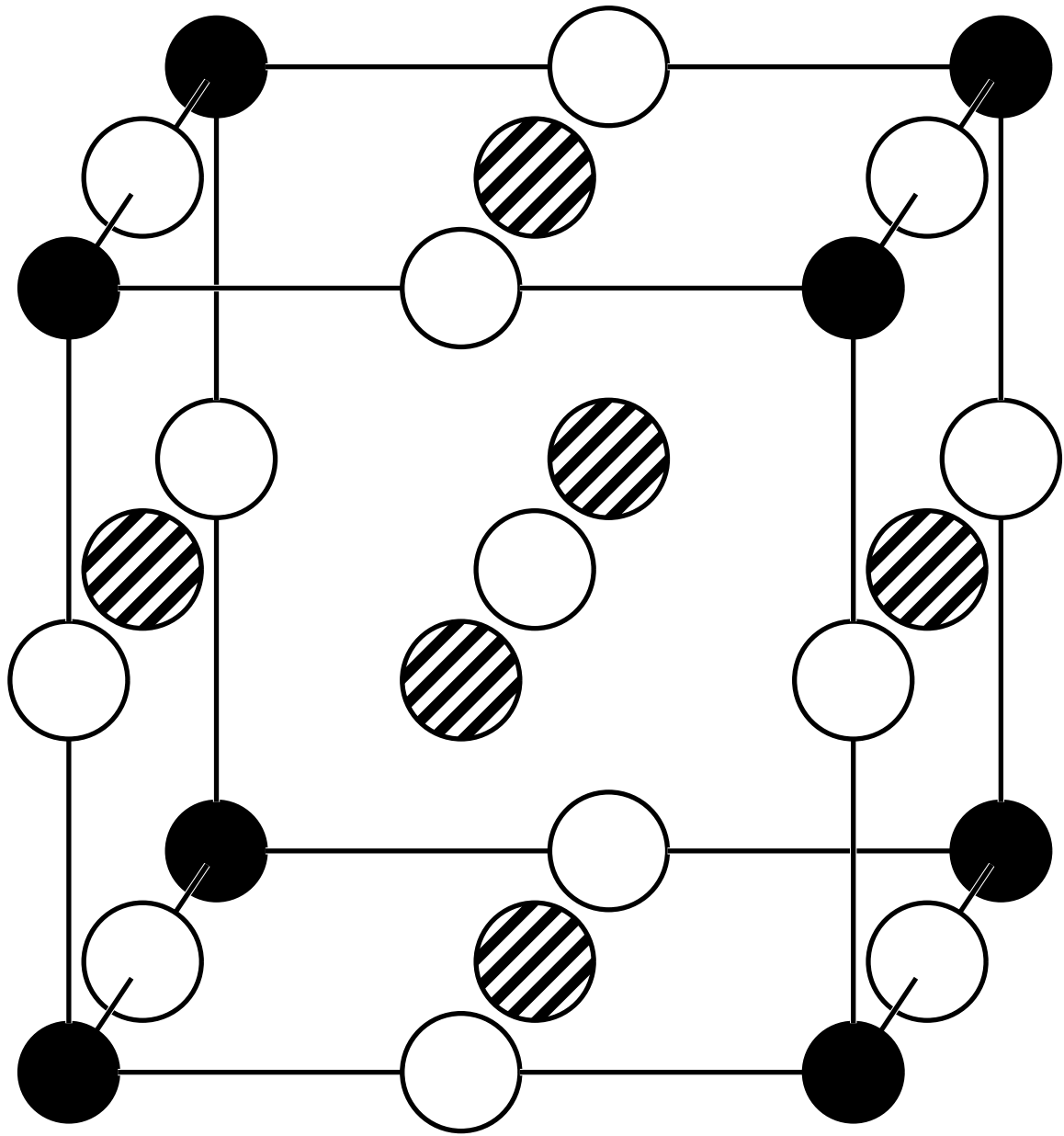
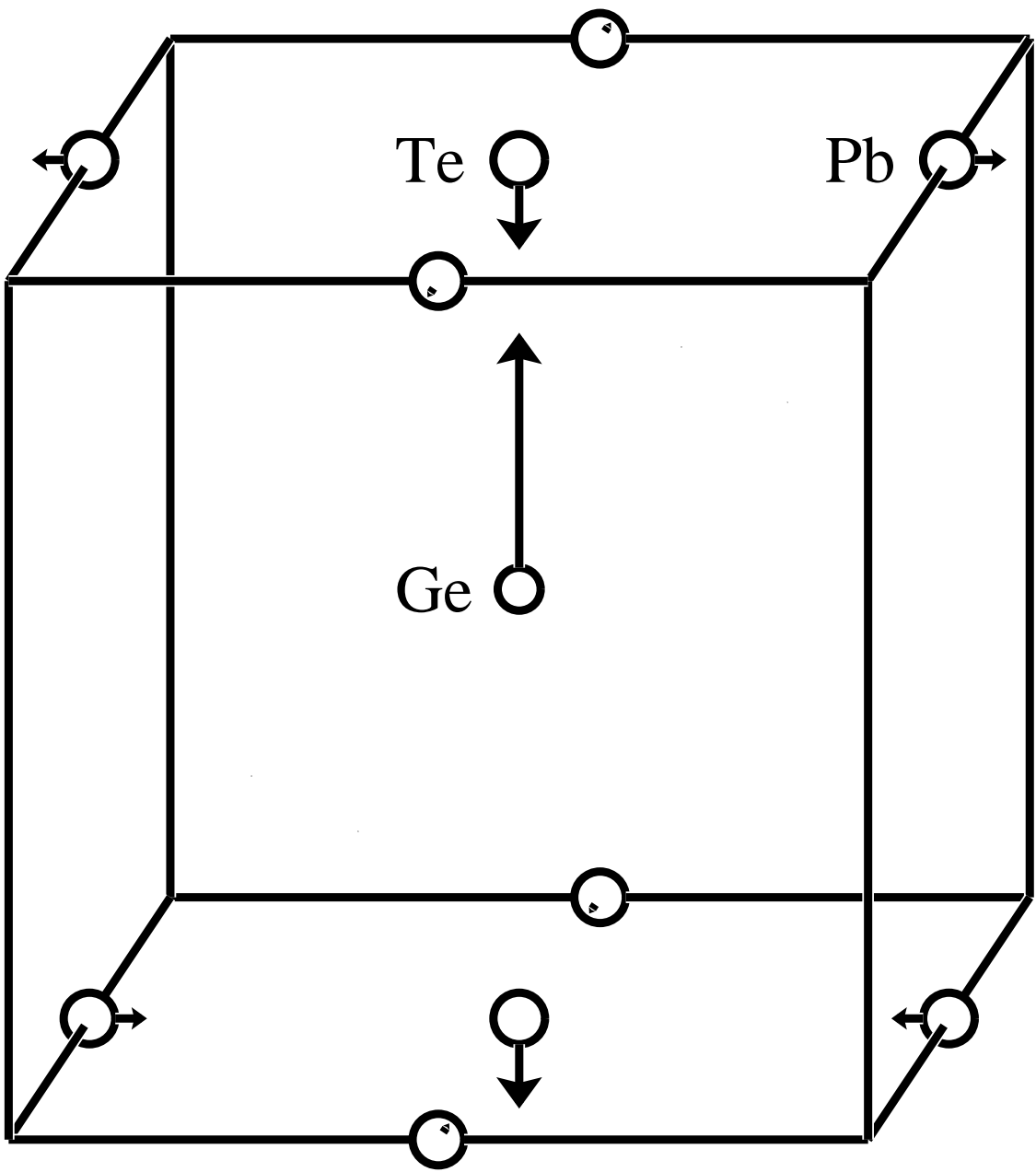
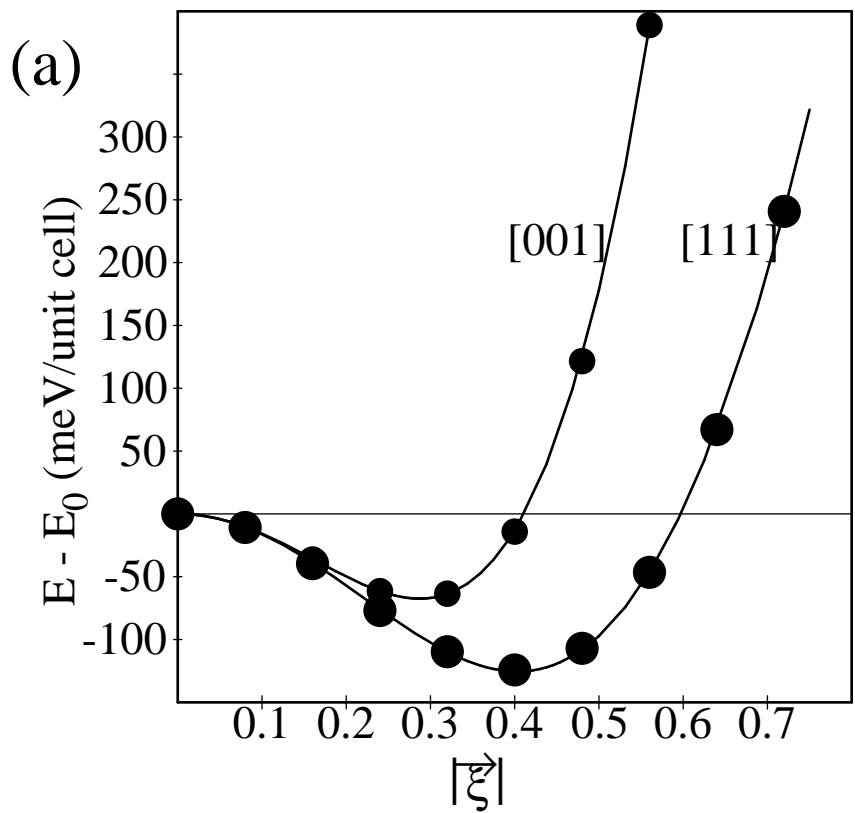
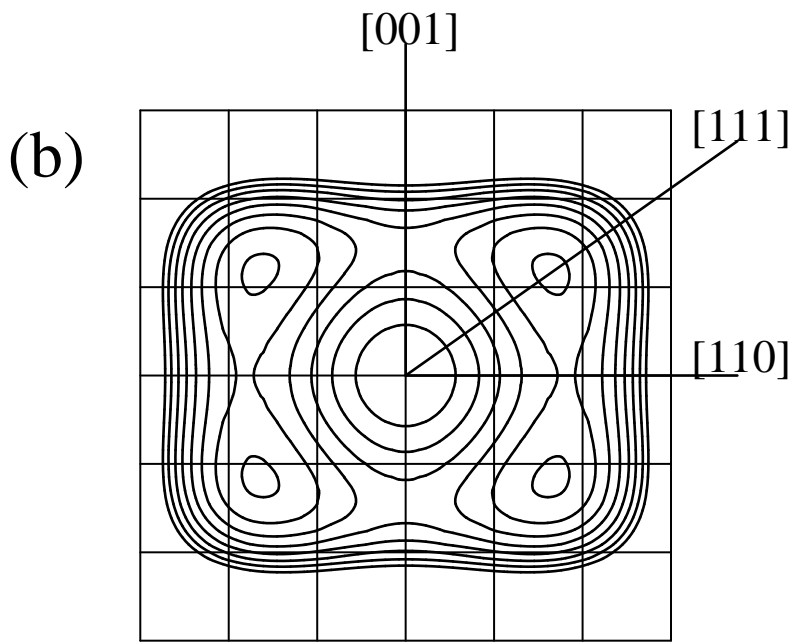
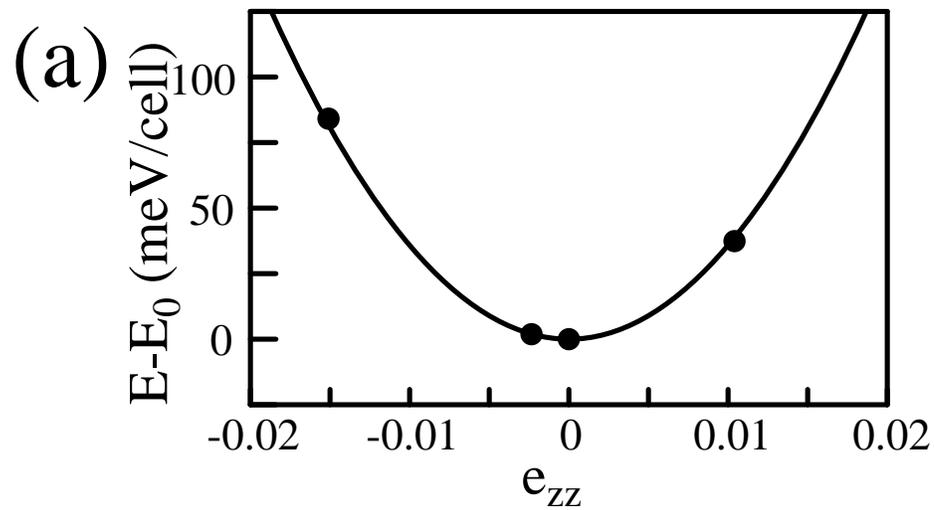
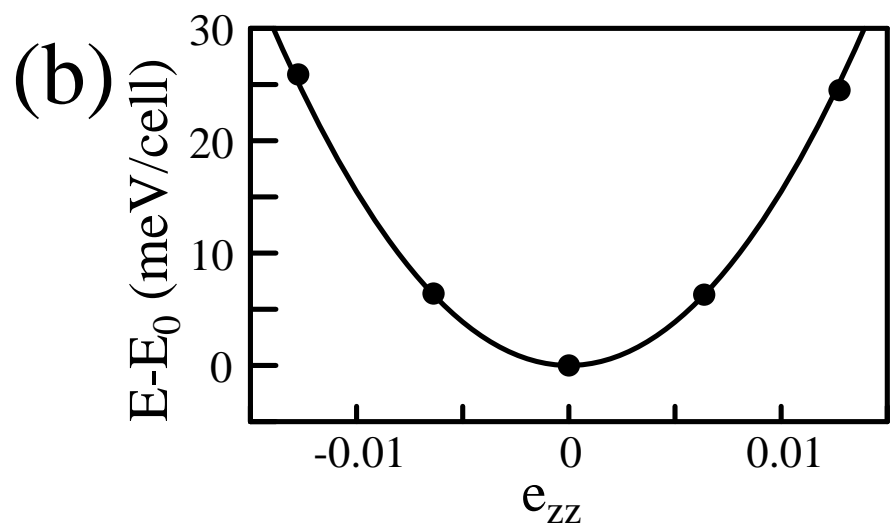
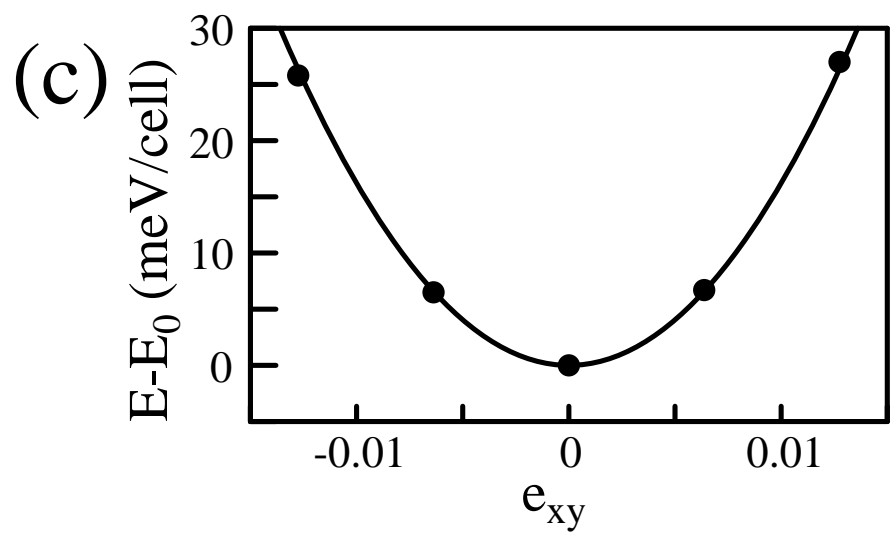
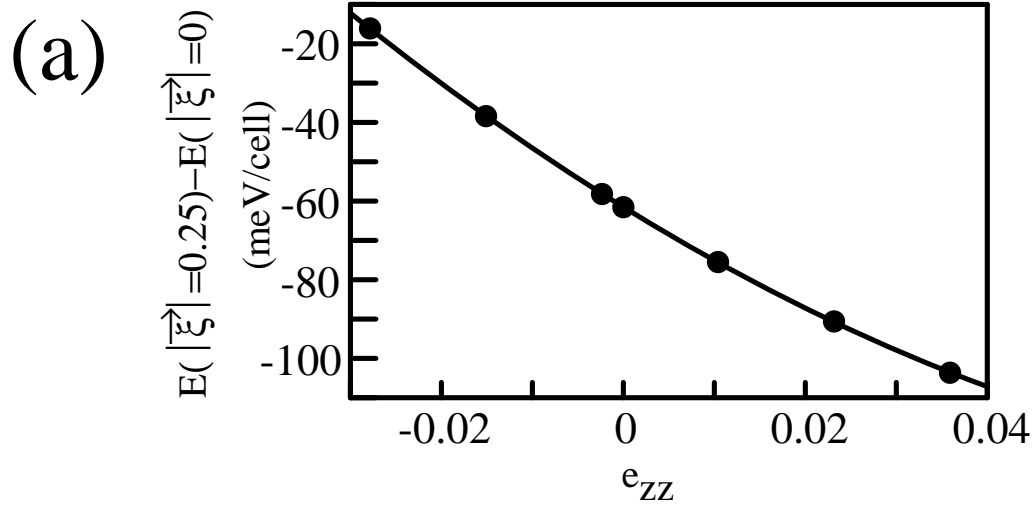
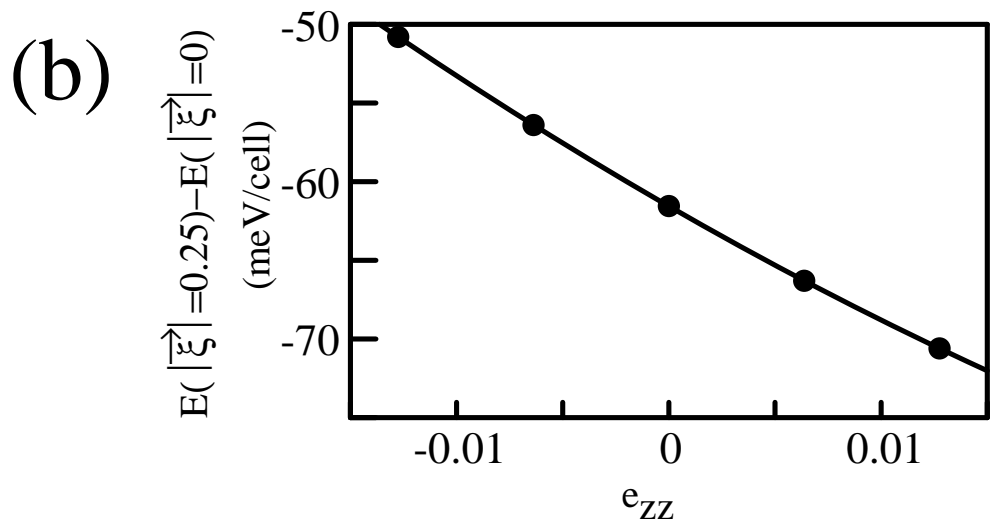
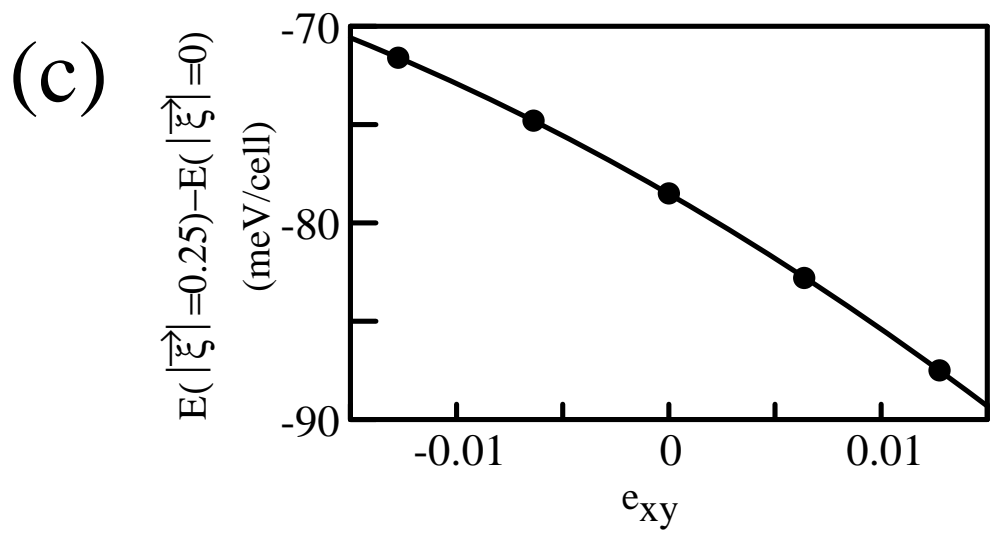
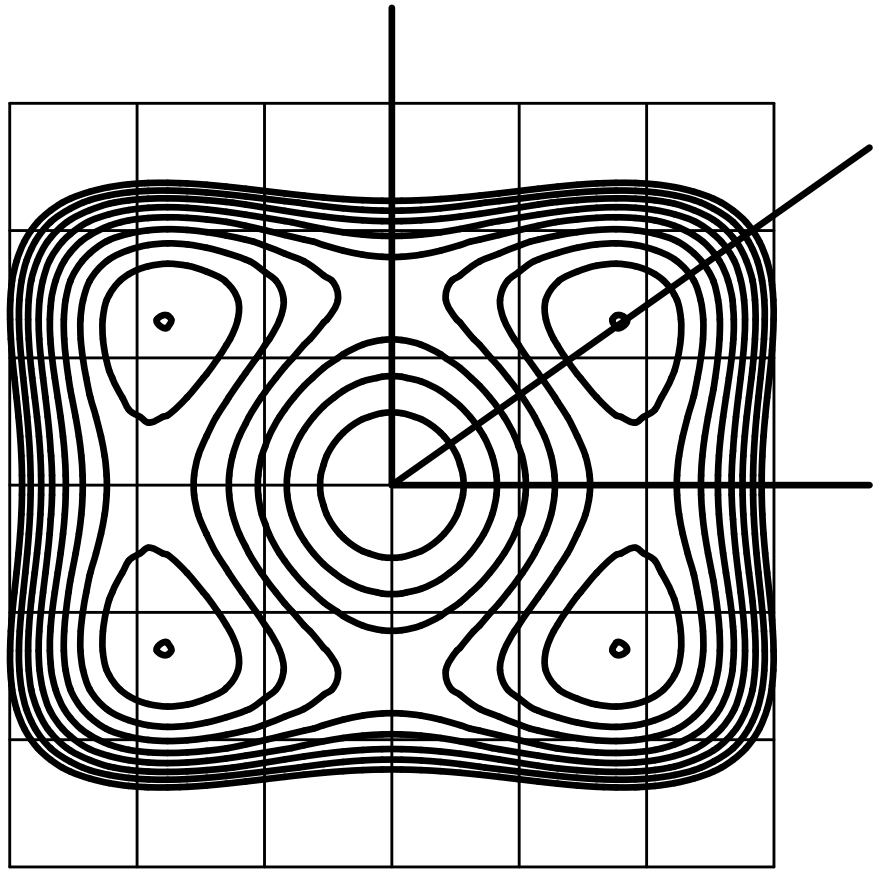
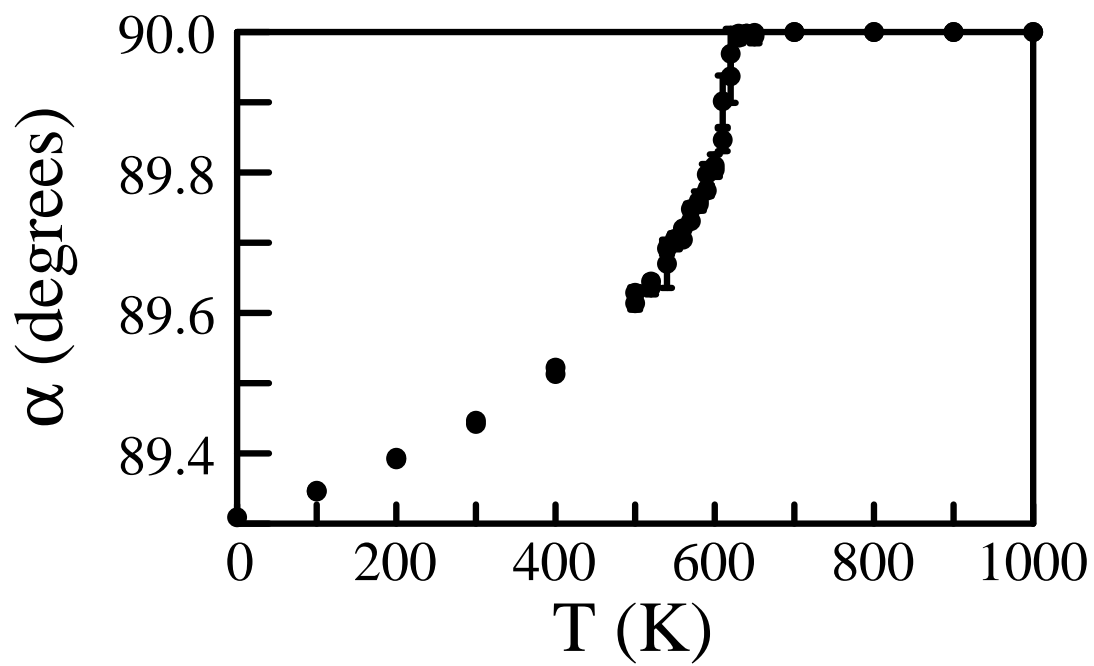
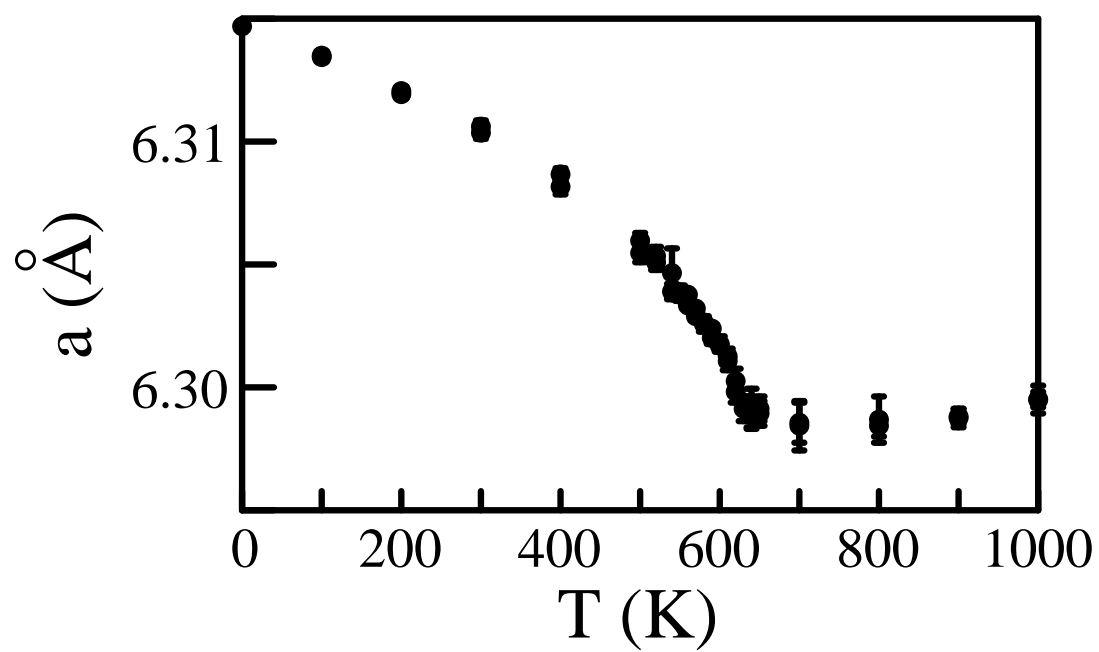
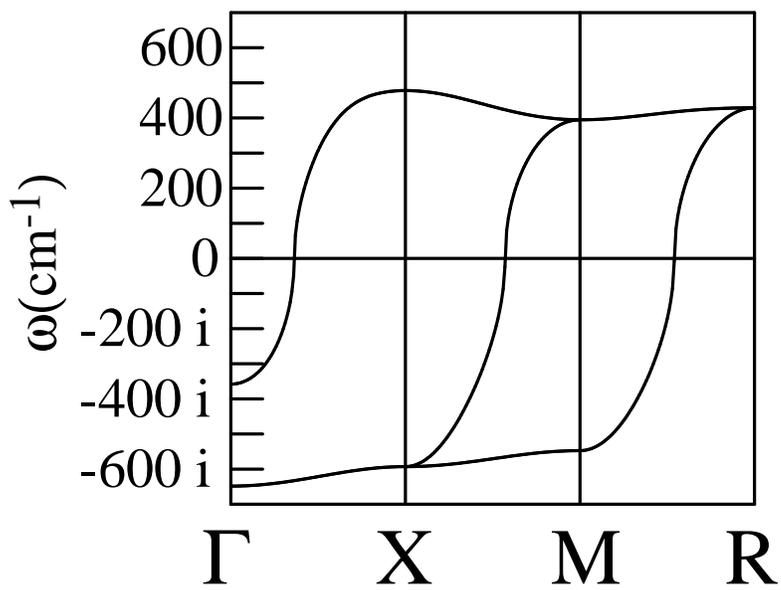
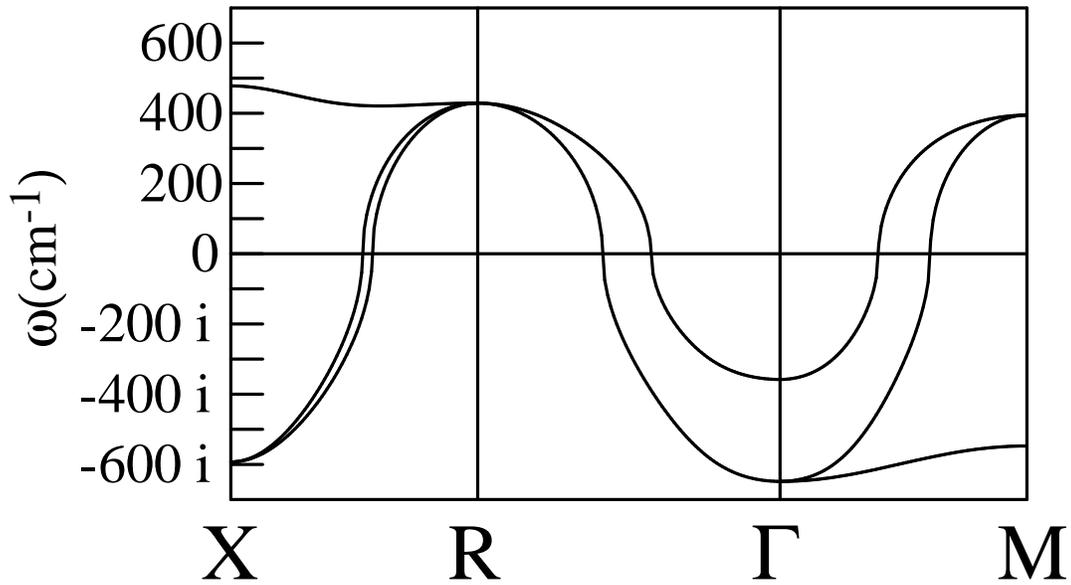
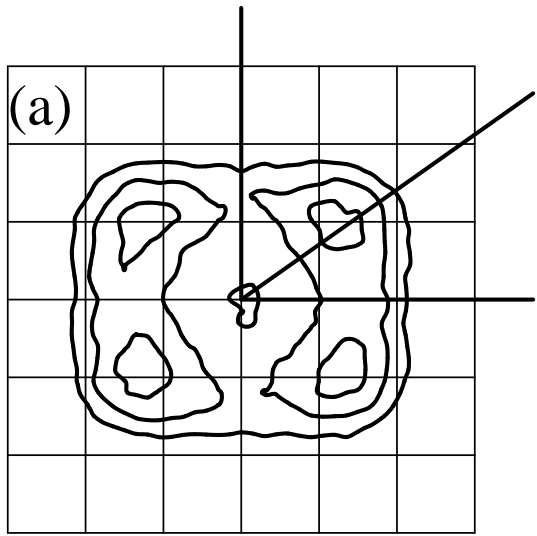
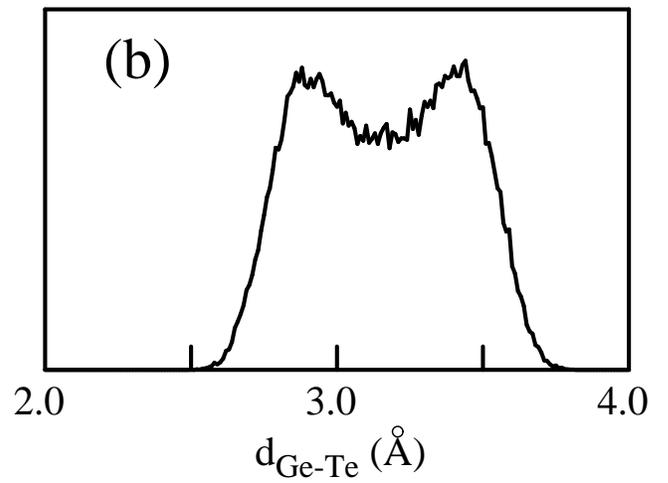
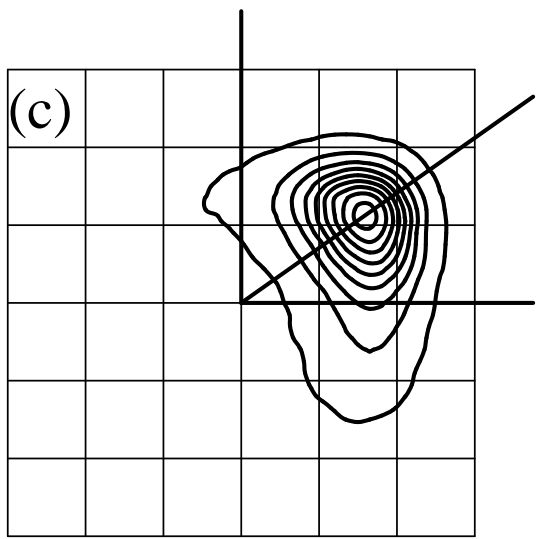
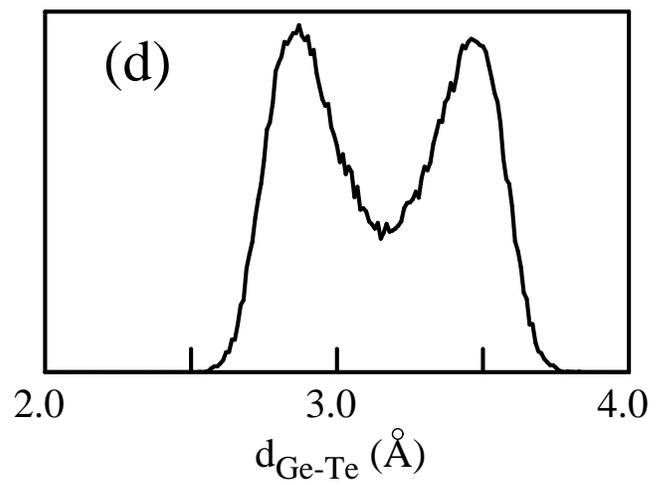
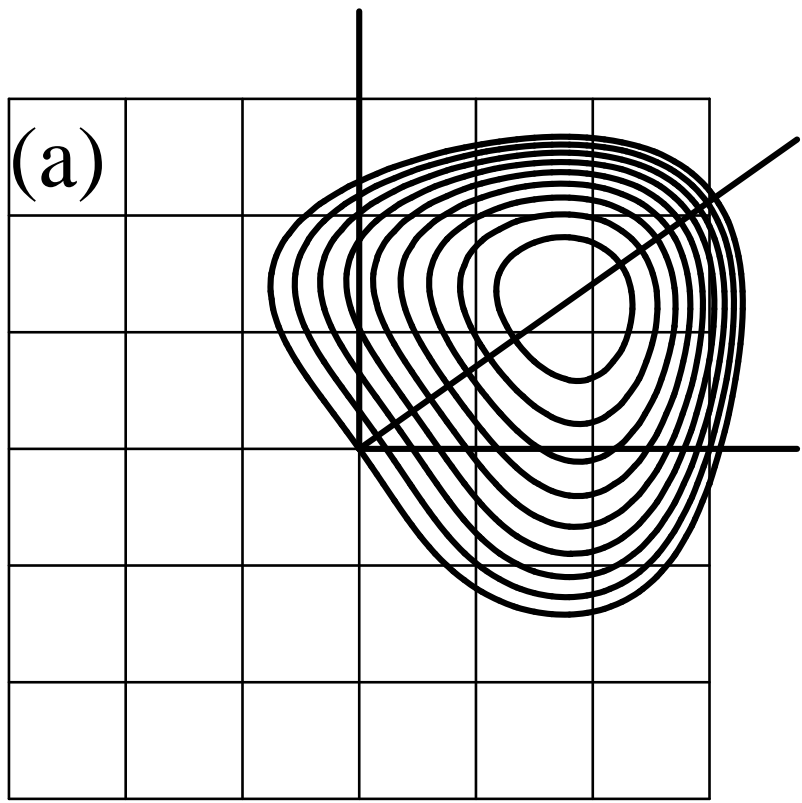
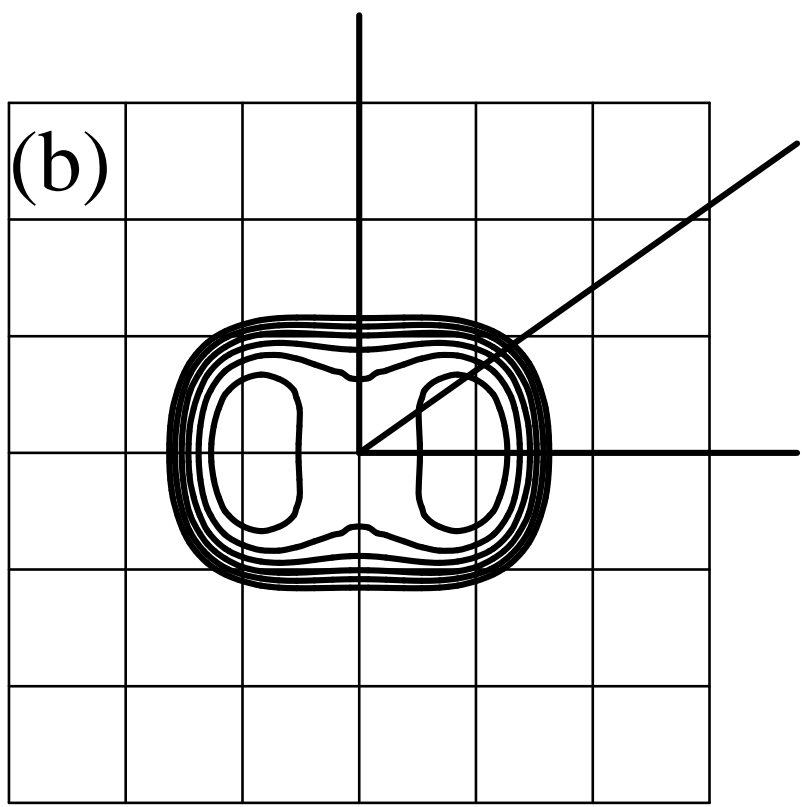
